\documentclass[11pt]{article}
\usepackage{natbib}
\usepackage{amsmath,amssymb,amsfonts,amsthm,graphicx,nicefrac,mathtools}

\usepackage[plain,noend]{algorithm2e}
\usepackage[english]{babel}
\usepackage[vmargin=1.18in,hmargin=1.1in]{geometry}

\usepackage{longtable}

\usepackage[T1]{fontenc}
\usepackage{bbm,mdframed}
\usepackage[dvipsnames]{xcolor}
\usepackage{xspace}
\usepackage{rotating,multirow}
\usepackage{graphics,tikz}
\usepackage{grffile}
\usepackage{subcaption, lscape}
\usepackage{epstopdf} 
\usepackage{makecell}

\newcommand{\figref}[1]{Figure~\ref{fig:#1}}

\newcommand{\secref}[1]{Section~\ref{sec:#1}}

\newcommand{\lemref}[1]{Lemma~\ref{lem:#1}}

\newcommand{\eqnref}[1]{\eqref{eqn:#1}}

\newcommand{\PP}[1]{\textnormal{Pr}\!\left({#1}\right)} 
\newcommand{\EE}[1]{\mathbb{E}\left[{#1}\right]} 

\newcommand{\PPst}[2]{\text{Pr}\!\left({#1}\ \middle| \ {#2}\right)} 

\def\R{\mathbb{R}}

\newcommand{\ignore}[1]{}

\newcommand{\cT}{\mathcal{T}}
\let\emptyset\varnothing

\newtheorem{theorem}{Theorem}

\theoremstyle{definition}
\newtheorem{definition}{Definition}
\theoremstyle{remark}

\newtheorem{lemma}{Lemma}

\usepackage{tikz}
\usetikzlibrary{arrows}
\usetikzlibrary{shapes}

\makeatletter
\renewcommand{\algocf@captiontext}[2]{#1\algocf@typo. \AlCapFnt{}#2} 
\def\@algocf@capt@plain{top}
\renewcommand{\algocf@makecaption}[2]{%
  \addtolength{\hsize}{\algomargin}%
  \sbox\@tempboxa{\algocf@captiontext{#1}{#2}}%
  \ifdim\wd\@tempboxa >\hsize
    \hskip .5\algomargin%
    \parbox[t]{\hsize}{\algocf@captiontext{#1}{#2}}
  \else%
    \global\@minipagefalse%
    \hbox to\hsize{\box\@tempboxa}
  \fi%
  \addtolength{\hsize}{-\algomargin}%
}
\makeatother

\makeatletter
\newcommand{\dotfrac}[2]{
\mathchoice
{\ooalign{$\genfrac{}{}{0pt}{0}{#1}{#2}$\cr\leavevmode\cleaders\hb@xt@ .22em{\hss $\displaystyle\cdot$\hss}\hfill\kern\z@\cr}}
{\ooalign{$\genfrac{}{}{0pt}{1}{#1}{#2}$\cr\leavevmode\cleaders\hb@xt@ .22em{\hss $\textstyle\cdot$\hss}\hfill\kern\z@\cr}}
{\ooalign{$\genfrac{}{}{0pt}{2}{#1}{#2}$\cr\leavevmode\cleaders\hb@xt@ .22em{\hss $\scriptstyle\cdot$\hss}\hfill\kern\z@\cr}}
{\ooalign{$\genfrac{}{}{0pt}{3}{#1}{#2}$\cr\leavevmode\cleaders\hb@xt@ .22em{\hss $\scriptscriptstyle\cdot$\hss}\hfill\kern\z@\cr}}
}
\makeatother

\newcommand{\nulls}{\mathcal{H}^0}

\newcommand{\fdr}{\textnormal{FDR}}

\newcommand{\Simes}{\textnormal{Simes}}

\newcommand{\One}[1]{{\bf{1}}\left({#1}\right)}

\def\H{\mathcal{H}}
\def\Parents{\mathrm{Par}}
\def\Depth{\mathrm{Depth}}
\def\Subgraph{\mathrm{Sub}}
\def\Children{\mathrm{Child}}
\def\Leaves{\mathcal{L}}
\def\Rejections{\mathcal{R}}
\def\False{\mathcal{V}}

\newcommand{\defn}{\ensuremath{:\, =}}

\newcommand{\Dset}{\ensuremath{\mathcal{D}}}

\newcommand{\Rplus}{\ensuremath{\R_+}}

\newcommand{\widgraph}[2]{\includegraphics[keepaspectratio,width=#1]{#2}}

\newcommand{\DAGGER}{\DAGGERSHORT$\;$}
\newcommand{\DAGGERSHORT}{\texttt{DAGGER}}
\newcommand{\LORD}{\texttt{LORD}$\;$}
\newcommand{\SCRDAG}{\texttt{SCR-DAG}$\;$}
\newcommand{\BHDAG}{\texttt{BH-DAG}$\;$}

\newcommand{\SHOLM}{\texttt{structured-Holm}$\;$}

\newcommand{\image}[1]{\raisebox{-0.5\height}{\includegraphics[width=4.7cm, height=3.5cm]{#1}}}

\date{\today}

\title{\DAGGERSHORT: a sequential algorithm for FDR control on DAGs}

\begin{document}

\author{
  Aaditya Ramdas$^{+}$, Jianbo Chen$^*$, Martin J. Wainwright$^{*\dagger}$, Michael I. Jordan$^{*\dagger}$\\
  Department of Statistics and Data Science$^+$,
  Carnegie Mellon\\
  Departments of Statistics$^*$ and EECS$^\dagger$,
  University of California, Berkeley\\
  \texttt{aramdas@stat.cmu.edu}, \texttt{jianbochen@berkeley.edu}\\
  \texttt{wainwrig,jordan@stat.berkeley.edu}
}








\maketitle

\begin{abstract}
We propose a linear-time, single-pass, top-down algorithm for multiple testing on directed
acyclic graphs (DAGs), where nodes represent hypotheses and edges
specify a partial ordering in which hypotheses must be tested.  The
procedure is guaranteed to reject a sub-DAG with bounded false
discovery rate (FDR) while satisfying the logical constraint that a
rejected node's parents must also be rejected. It is designed for
sequential testing settings, when the DAG structure is known a priori,
but the \mbox{$p$-values} are obtained selectively (such as in a
sequence of experiments), but the algorithm is also applicable in
non-sequential settings when all $p$-values can be calculated in
advance (such as variable/model selection). Our \DAGGER algorithm,
shorthand for Greedily Evolving Rejections on DAGs, provably controls the false
discovery rate under
independence, positive dependence or arbitrary dependence of the
$p$-values. 
The \DAGGER procedure
specializes to known algorithms in the special cases of trees and line
graphs, and simplifies to the classical Benjamini-Hochberg procedure
when the DAG has no edges. We explore the empirical performance of
\DAGGER using simulations, as well as a real dataset corresponding to
a gene ontology, showing favorable performance in terms of time and
power.
\end{abstract}


\section{Introduction}
\label{sec:intro}

Considerable effort in the field of multiple testing is devoted to
incorporating prior knowledge and structural constraints into the
testing procedure, in the form of covariates, weights, groupings of hypotheses,
knowledge of dependence structures, and so on. In this paper, we consider a general setting
in which nodes representing hypotheses are organized in a directed
acyclic graph (DAG), where the directed edges encode a partial order
in which the hypotheses must be tested.  We present a sequential top-down
algorithm called \DAGGER that is designed to efficiently test hypotheses
on such a DAG while providing false discovery rate (FDR) control.

Research on multiple testing on trees and DAGs is
  relatively recent.  \citet{meinshausen2008hierarchical} derived a
  method for familywise error rate (FWER) control on trees for
  hierarchical testing of variable importance, and
  ~\citet{yekutieli2008hierarchical} presented a method for FDR
  control on trees (under independence) for multiresolution testing of
  quantitive trait loci.  \citet{goeman2008multiple} proposed an
  algorithm for FWER control on a DAG, applied to the gene ontology
  graph. This earlier work left open a fourth natural combination,
  which is FDR control on DAGs. Progress was made by
  \citet{lynch2014control}, but the resulting algorithms had several
  limitations: they were quadratic-time, multi-pass algorithms that
  cannot handle arbitrary dependence between p-values. In the current
  paper we move beyond those limitations---we present a linear-time,
  single-pass algorithm that can handle any form of dependence.
  Moreover, based on experimental results, we find that our algorithm
  is more powerful than Lynch's method.

The aforementioned gene ontology graph is an intersection DAG, where
every node corresponds to an intersection of elementary hypotheses,
and the intersection set of a parent node is a superset of the union
of its children. Hence, the set of non-null hypotheses must satisfy the
\begin{quote}
Strong Heredity Principle: all parents of a non-null node are also
 non-null.
\end{quote}
Logical constraints that hold in reality need not necessarily hold for
the outputs of procedures such as the Benjamini-Hochberg or Bonferroni 
methods that control FDR or FWER. In order to impose such a logical structure on
the output of a multiple-testing procedure, we must explicitly insist
that the rejected hypotheses obey the constraints:
\begin{itemize} 
\addtolength{\itemindent}{0.6cm}%
\item[(C1).] The rejected nodes must form a DAG (a
  subgraph of the original DAG).
\item[(C2).] All parents of a rejected node must themselves be
  rejected.
\end{itemize}
Of course condition (C2) implies condition (C1), but not vice versa,
and we make the difference explicit to provide intuition.

While some multiple-testing problems are non-sequential in nature, meaning that
all the $p$-values are available at once, we wish to consider the more
general problem of sequential multiple testing on arbitrary DAGs (that
need not be of the intersection type), where the $p$-values are not
available all at once, but must be obtained in sequence, possibly at
some cost.
To summarize, let us isolate two distinct motivations for algorithms
on DAGs :
\begin{enumerate}
  \addtolength{\itemindent}{0.6cm}%
\item[(M1).] A DAG can represent a partial ordering over a collection
  of hypotheses to be tested, where we first conduct tests at all the
  roots, and then proceed to test hypotheses selectively down the DAG depending on their outcomes.  We can save resources by
  exploiting constraints (C1) and (C2), by not testing descendants of
  a node that is not rejected, while further exploring the subgraph of
  nodes that are rejected. At the end of the adaptive exploration
  process, we would like a guarantee that not too many rejected nodes
  are false discoveries.
\item[(M2).] A DAG can represent a structural constraint that one
  might desire for interpretability of the rejected set, even when the
  $p$-values at all nodes can be calculated offline from existing data
  and there is no sequential component to the problem. We encounter
  this setting when dealing with a gene ontology DAG later in this
  paper, which is an intersection DAG. Here, logical coherence would
  require that the underlying truth must satisfy both (C1) and (C2).
\end{enumerate}

Under motivation (M1), the algorithm must be a single-pass and
top-down algorithm, due to the nature of the investigative
process. However, under motivation (M2), any top-down or bottom-up or
combined algorithm would suffice, as long as the set of rejected nodes
satisfies requirements (C1) and (C2).  We remark that the Strong
Hierarchy Principle represents a set of one-way constraints: a two-way
constraint would further require that every rejected parent has at
least one rejected child.  Note that two-way constraints need not
necessarily make sense under motivation (M1); moreover, scientists
having motivation (M2) may be interested in group-level findings even
without individual-level identification.  Accordingly, our analysis
here focuses exclusively on one-way constraints.

In this paper, we design a top-down sequentially rejective algorithm called \DAGGER
 for FDR control on DAGs.  It was
designed based on motivation (M1), but may naturally also be applied
in situation (M2). Specifically, \DAGGER can be run in a
truly sequential fashion since it does not access child $p$-values
when deciding whether to reject a parent or not. To the
  best of our knowledge, \DAGGER is the first algorithm
  that, in either setting (M1) or setting (M2) is able to explicitly
  take into account the structure of the DAG, and can provably control
  FDR on general or intersection DAGs under a variety of dependence
  assumptions.

There are several special structures that are
of relevance: a tree (with one root, $L$ leaves, and each non-root
node having at exactly one parent), a forest (a disjoint collection of
trees), a line (with one root, one leaf, and each internal node having
one parent and one child), and the empty graph (which corresponds to
the standard unconstrained multiple testing problem). When applied to
trees/forests, \DAGGER specializes to the algorithm for
FDR control on trees/forests introduced by \citet{lynch2016control};
this algorithm in turn specializes to the algorithm for FDR control on a line by
\citet{lynch2016procedures}, which in turn simplifies to the procedure
by \cite{BH95} on a graph without edges.  

\citet{lynch2014control} focuses on DAGs, and designs two
  new algorithms for DAGs, called BH-DAG and SCR-DAG.
  Although these provide FDR control on DAGs, they do not resolve the
  problems that we have captured in motivations (M1)
  and (M2). To elaborate, both methods are multi-pass and
  non-sequential, and hence cannot be used for adaptive
  experimentation under (M1). Further, they cannot deal with the arbitrary
  dependence between hypotheses under (M2) that arises, for example, in multiple
  testing on the gene ontology DAG.  

There has been some work on online multiple testing for a
  fully ordered set of hypotheses.  \cite{rosenbaum2008testing}
  proposed a sequential algorithm for FWER control.  In parallel, \cite{FS08}
  proposed online FDR algorithms called alpha-investing, which
  have been further studied and generalized by other
  researchers~\citep{aharoni2014generalized,JM16,RYWJ17,
    SAFFRON}. One may, somewhat arbitrarily, convert a partial ordering
  into a full ordering and then apply one of the above algorithms. Doing so ignores
  the structure of the partial ordering, however, leading to suboptimal performance
  in practice, as confirmed by our experiments. Further, none of these online FDR algorithms,
  with the exception of LORD, work under arbitrary dependence. 

  Fully-ordered testing algorithms have
  been derived for model selection by
  \cite{g2013false}, \cite{knockoffs} and
  \cite{li2017accumulation}. The STAR algorithm of \cite{lei17star},
  and its reversed counterpart in \cite{katsevich2018towards}, extend these to the setting of partial orderings by utilizing user interaction. 
  However, all these algorithms are inherently non-sequential (requiring all p-values up front) 
  and also need independence of $p$-values, rendering them inapplicable under both
  motivations (M1) and (M2).

\section{Problem setup and semantics}
\label{sec:semantics}

A directed acyclic graph
(DAG) is a graph with directed edges and no directed cycles.
We consider a DAG with $N$ nodes, each
representing one hypothesis associated with a single $p$-value.  The
hypothesis at a node may either correspond to (a) the intersection
hypothesis of its children; or (b) a separate, possibly unrelated
hypothesis.  A DAG where every node corresponds to the first type is
called an intersection DAG. Our algorithms will apply to
intersection DAGs as well as more general DAGs.

Let us call the $N$ hypotheses $H_1,\ldots,H_N$, and the set of all
hypotheses as $\H$, and these are associated with corresponding
$p$-values $P_1,\ldots,P_N$. Naturally, when any of the hypotheses is a
true null, we assume that its $p$-value is super-uniform, meaning that
it satisfies the inequality
\begin{align}
  \label{eqn:super-uniform}
\PP{P \leq t} \leq t \text{~ for all ~} t \in [0,1].
\end{align}

We use the terms parent and child in the natural way: if $A \rightarrow B$ is an edge, then $A$ is
one of the parents of $B$, and $B$ is one of the children of
$A$. Nodes without parents are called roots, while nodes without
children are called leaves, while internal nodes have both
parents and children.  Analogously, the ancestors of $B$
refer to all nodes, including its parents, that may follow a directed
path that leads to $B$, while the descendants of $A$ refer to all
nodes, including its children, that it may reach using a directed
path.  For the purposes of visualization, we imagine roots at the top,
leaves at the bottom and all edges pointing downwards; we then use the
term top-down to refer to calculations performed from roots to
leaves, and bottom-up to mean the opposite.

We define the depth of a node $a$, denoted $\Depth(a)$, in a top-down
manner as the length of the longest path from any root to the node,
plus one. All roots are initialized to have depth one. Any node that
has all parents being roots has depth two, and so on recursively. Then,
the depth of a node is always one larger than the maximum depth of any
of its parents.  More formally, using $\Parents(a)$ to refer to the
 parents of node $a$, we have
\begin{align*}
\Depth(a) = 1 + \max_{b \in \Parents(a)} \Depth(a).
\end{align*}
In other words, a node's depth is the length of the longest possible
path one could take from a root to the node. Let $D$ denote the
maximum depth of any node in the DAG and let $\Leaves$ denote the set
of all leaves. Note that all roots have depth one, but not all leaves
have depth $D$---there might be leaves at depths smaller than $d$, and
in fact a node that is isolated from the graph is both a root and a
leaf at depth one. Let $\H_d$ denote the set of all hypotheses at depth
$d$, and let us arbitrarily name these as $H_{d,1},\dots,H_{d,|\H_d|}$
with associated $p$-values $P_{d,1},\dots,P_{d,|\H_d|}$. Let
$\H_{1:d}$ denote the set of all nodes with depth $\leq d$. Then
$\{\H_d\}_{d=1}^D$ is a partition of the hypotheses that naturally
satisfy $\H_d \cup \H_{1:d-1} = \H_{1:d}$ and also $\H_{1:D} = \H$.

\begin{figure}[h!]
  \centering \widgraph{0.5\linewidth}{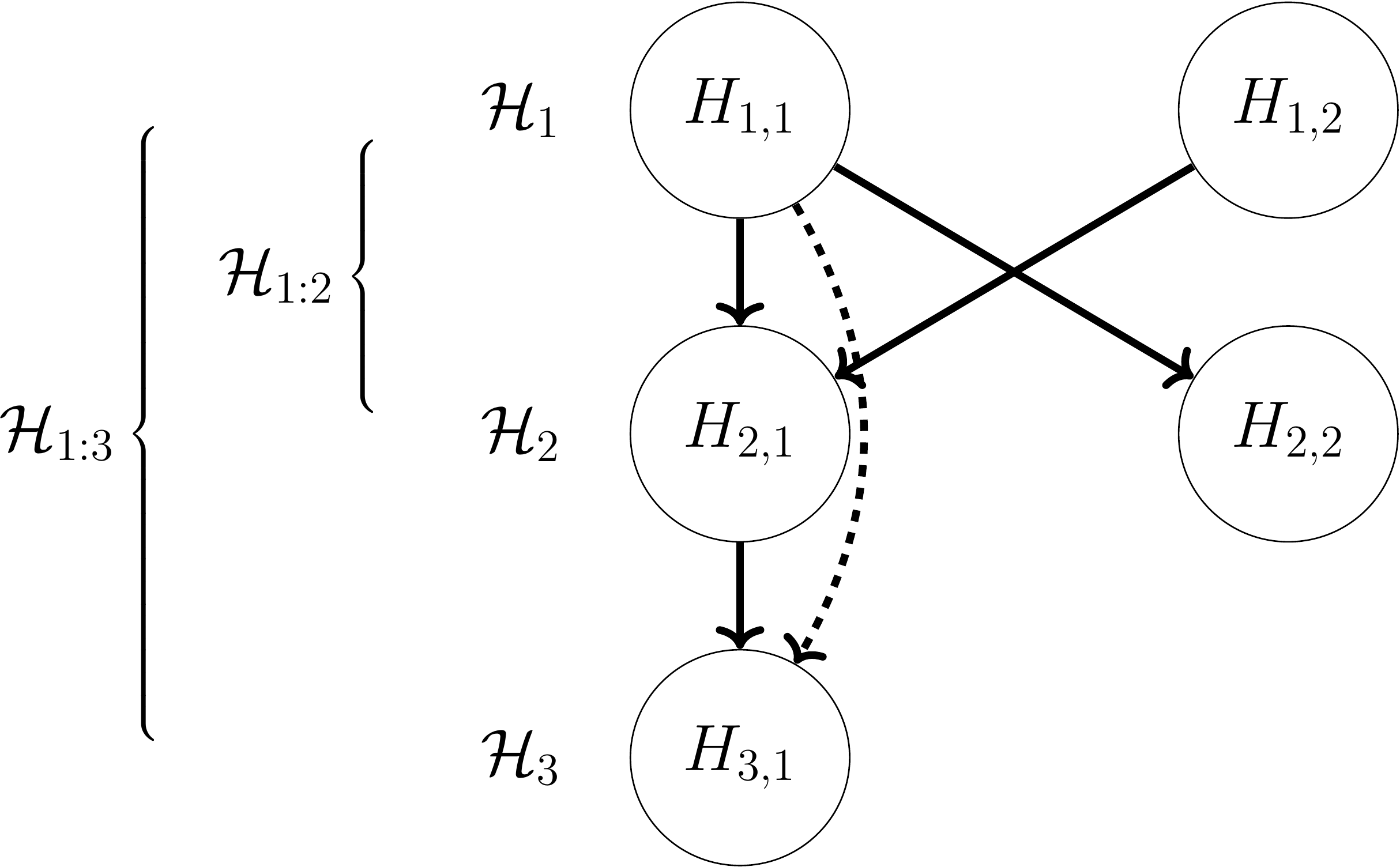}
  \caption{An example of a DAG with $D=3$. The roots
    $H_{1,1}, H_{1,2}$ are necessarily at depth one, while the leaves
    $H_{2,2},H_{3,1}$ may occur at different depths. We do not allow
    any edge such as the one from $H_{1,1}$ to $H_{3,1}$ (dotted for
    clarity), since it does not encode any new constraints.}
\label{fig:layers}
\end{figure}

Note that we sometimes refer to nodes in the graph with the subscripts
$a \equiv (d,i)$.  Consequently, if we explicitly refer to an
arbitrary hypothesis while making its depth explicit, we may use the
label $H_{d,i}$, but if its depth is implicit then we may simply refer
to the node as $H_a$. See \figref{layers} for an illustrative example.

\subsection{Testing protocol and goal}

Recall that we assume the scientist creates, or knows, the structure
of the DAG in advance to performing any tests. The goal of the
scientist is to design testing levels $\alpha_{d,i}$ at which to test
each $P_{d,i}$, so that the false discovery rate (FDR) over the entire
DAG is controlled at a predefined level $\alpha$. The process is
inherently constrained by the DAG to be sequential in nature---before
testing a node, the scientist must have already tested and rejected
all of its parents. In other words, the testing
proceeds in $D$ steps:
\begin{enumerate}
\item Test all root hypotheses at some predefined constant levels
  $\{\alpha_{1,i}\}_{i \in \H_1}$, and let $\Rejections_1$ denote the
  corresponding set of rejections.
\item For each depth $d=2,\ldots,D$, repeat the following: test all
  hypotheses at depth $d$ at levels $\{\alpha_{d,i}\}_{i \in \H_d}$,
  and let $\Rejections_d$ denote the corresponding set of rejections.
\end{enumerate}
Let $\Rejections_{1:d} = \bigcup_{k=1}^{d} \Rejections_k$ be the total
set of rejections up to depth $d$. Then, we formally require that the
level $\alpha_{d,i}$ at which $P_{d,i}$ is tested must be solely a
function of $\Rejections_{1:d-1}$ and the $p$-values in $\H_d$.

Denote the set of discoveries as the rejected set
$\Rejections = \bigcup_{d=1}^D \Rejections_d$, of size $R =
|\Rejections|$. If $\nulls$ is the unknown set of true null
hypotheses, then let the set of false discoveries be $\False = \nulls
\cap \Rejections$, of size $V = |\False|$. Clearly, both $\Rejections$
and $\False$ are random variables, and we define the false discovery
rate as their expected ratio, that is $\fdr=\EE{V/R}$. The ratio $0/0$ is always
  interpreted as $0$.

Since we only test nodes whose parents have been rejected, so the set of
nodes that is tested is itself random and
unknown a priori.  Intuitively this model mimics the adaptive process of
science, allowing for the subset of hypotheses to be tested in the
future to depend hypotheses are currently being tested, which in turn
is a function of which hypotheses have been tested and rejected in the
past.  We wish to design algorithms to set the testing levels
$\alpha_i$ in an adaptive manner, so that no matter which nodes are
null and non-null, and no matter the randomness in the corresponding
$p$-values, the false discovery rate over the whole DAG is
controlled. It turns out the algorithm design, as well as power,
depend on what assumptions one is willing to impose on the dependence
structure between the various $p$-values, an issue which we now
discuss.


\subsection{Positive dependence}

In this paper, we analyze four possible settings of dependence between
the $p$-values, which are defined in terms of sets and functions that
are nonincreasing with respect to a partial ordering.  For a pair of
vectors $x, y \in [0,1]^K$, we use the notation $x \preceq y$ to
denote the partial ordering defined by the orthant cone: i.e., we have
\mbox{$x \preceq y$} if and only if \mbox{$x_i \leq y_i$} for all
\mbox{$i \in \{1, \dots, K\}$.}

A function \mbox{$f: [0,1]^K \mapsto \Rplus$} is nonincreasing
  with respect to the orthant ordering if \mbox{$x \preceq y$}
implies \mbox{$f(x) \geq f(y)$.}  Similarly, a subset $\Dset$ of
$[0,1]^K$ is nondecreasing with respect to the orthant ordering
if \mbox{$x \in \Dset$} implies \mbox{$y \in \Dset$} for all \mbox{$y
  \succeq x$.}  We now define a notion called positive regression dependence on a subset, or positive dependence for short, for a vector of $p$-values $P$.
\begin{definition}[positive regression dependence on a subset]
\label{ass:PRDS}
We say that the vector $P$ is positively dependent if
for any  $i \in \nulls$ and nondecreasing set $ \Dset \subseteq
[0,1]^n$, the function $t~\mapsto~\PPst{P\in \Dset}{P_{i} \leq t}$
is nondecreasing on the interval $(0,1]$.
\end{definition}
When the condition is met for a particular null $P_{i}$, we
say that $P$ is positively depdendent on $P_{i}$.  Note that the original positive
regression dependence assumption~\citep{lehmann1966some} and that of positive regression dependence on a subset~\citep{BY01}, both had $P_{i} = t$ instead of $P_{i}
\leq t$ in the definition, but one can prove that both conditions are
essentially equivalent.
Positive dependence holds trivially if the $p$-values are independent,
but also allows for some amount of positive dependence.  For example, let $Z =
(Z_1,\ldots,Z_n)$ be a multivariate Gaussian vector with covariance
matrix $\Sigma$; the null components correspond to Gaussian variables
with zero mean.  Letting $\Phi$ be the cumulative distribution
function of a standard Gaussian, the vector
$P=(\Phi(Z_1),\dots,\Phi(Z_n))$ is positive regression dependent on $P_i$ for every null index
$i$ if and only if all entries of the covariance matrix $\Sigma$ are
non-negative.  See~\citet{BY01} for additional examples of this
type. We note that Definition~\ref{ass:PRDS} is closely related to the
assumption of log-supermodularity, or equivalently multivariate total
positivity of order two, as studied by \citet{karlin1980classes}.
Since log-supermodularity is known to imply positive regression dependence, our results
also hold immediately under the former condition.


\subsection{Arbitrary dependence and reshaping}

In order to deal with arbitrarily dependent $p$-values, we first need
to define reshaping functions $\beta$, as introduced
by~\citet{blanchard2008two}.

\begin{definition}[Reshaping]
Let $\cT$ denote the set of all probability measures $\tau$ with
domain $\R^+$. Given any $\tau \in \cT$, the reshaping function $\beta
= \beta_\tau$ is given by
\begin{align*}
\beta_\tau(r) = \int_{0}^r x d\tau(x).
\end{align*}
We let $\beta(\cT) \defn \{\beta_\tau \mid \tau \in \cT\}$ refer to
the set of all such reshaping functions.
\end{definition}
\noindent Note that one may always choose $\beta(r)=r$ if one wishes
to avoid reshaping, as is done when dealing with independent or
positively dependent $p$-values. We often drop the subscript $\tau$
for ease of notation.

When no assumptions are made about the joint distribution of
$p$-values, FDR-controlling procedures must generally be guarded while
proclaiming a discovery. Noting that $\beta(r) \leq r$ for any
$\tau \in \cT$ by construction, the function $\beta$ is guaranteed to
reshape thresholds by lowering them, rendering the associated
procedure more conservative.  Indeed, the argument $r$ has been used
suggestively, since $\beta$ conservatively undercounts the number of
rejections. Effectively, \citet{BY01} use the normalized discrete measure
assigning mass $\propto 1/k$ to each positive integer $k \in \{1,\dots,K\}$ as
a specific choice of $\tau$, leading to the reshaping function
$\beta_{BY}(r) = \frac{r}{\sum_{k=1}^K \frac{1}{k}}$ that underlies
the Benjamini-Yekutieli procedure.  
  See~\citet{ramdas2017unified} for further
discussion on the uses and interpretation of reshaping, and
associated references.


\subsection{The Simes' $p$-value}

Given a subset $\mathcal{S}$ of hypotheses $H_1,\ldots,H_S$ and
associated $p$-values $P_1,\ldots,P_S$, let $P_{(1)},\ldots,P_{(S)}$
denotes the sequence of ordered $p$-values. Then, the generalized
Simes' $p$-value is given by $\Simes(\mathcal{S}) \defn \min_{(k)}
\frac{P_{(k)} \cdot S}{k}$ under independence or positive dependence
of the underlying $p$-values, and $\Simes(\mathcal{S}) \defn
\min_{(k)} \frac{P_{(k)} \cdot S}{\beta(k)}$ under arbitrary
dependence.  It is well known that if $\mathcal{S}$ is null, meaning
that it consists only of true null hypotheses, then
$\Simes(\mathcal{S})$ is a bonafide $p$-value for the intersection
hypothesis \mbox{$H_1 \cap \dots \cap H_S$.}  More precisely, this
statement means that under the global null hypothesis for subset
$\mathcal{S}$, we have $\PP{\Simes(\mathcal{S}) \leq t} \leq t$ for
all $t \in [0,1]$.
For proofs of the above claims, see the
papers by~\citet{Simes1986improved}, \citet{BY01}, \citet{blanchard2008two}
and \citet{ramdas2017unified}.

The Simes' $p$-value will be of special interest for intersection
DAGs, because they yield bonafide $p$-values under positive
dependence, even without reshaping. We remark that we may use other
methods for combining $p$-values within a group to test the
intersection null for that group, such as Fisher's or Rosenthal's, as
long as the appropriate independence assumptions within the group are
satisfied, and as long as we use reshaping to guard against arbitrary
dependence among the resulting group $p$-values.


\section{An algorithm for general DAGs}
\label{sec:alg1}

The settings of dependence that can be handled by our
algorithm are as follows:
\begin{enumerate}
\item[(D1).] For arbitrary and intersection DAGs: all $p$-values are
  independent.
\item[(D2).] For arbitrary and intersection DAGs: all $p$-values are positively dependent.
\item[(D3).] For arbitrary and intersection DAGs: all $p$-values are
  arbitrarily dependent.
\item[(D4).] For intersection DAGs: the leaf $p$-values are positively dependent, and
  other $p$-values are formed using Simes' procedure.

\item[(D5).] For intersection DAGs: the leaf $p$-values are independent, and
  other $p$-values are formed using Fisher's procedure, or the procedure
  presented by \citet{stouffer1949american}, commonly called Rosenthal's method.

\item[(D6).] For intersection DAGs: the leaf $p$-values are
  arbitrarily dependent, and other $p$-values are formed using the
  procedures by \citet{ruschendorf1982random},
  \citet{vovk2012combining} or \citet{ruger1978maximale}.
\end{enumerate}

In order to describe our algorithm, we need some further DAG-related
notation. The descendants of a node $a$ is the set of nodes that can
be reached from $a$ along a path of directed edges. We use
$\Subgraph(a)$ to denote the subgraph formed by taking $a$ as the
root, along with all descendants of node $a$.  The effective
  number of nodes and effective number of leaves in
$\Subgraph(a)$, denoted by $m_a$ and $\ell_a$ respectively, are
defined as follows. We calculate $\ell_a, m_a$ for each node $a$ in a
bottom-up fashion: we first instantiate each leaf node $a \in \Leaves$
with the value $m_a = \ell_a = 1$, and then proceed up the tree, from
leaves to roots, recursively calculating
\begin{align}
  \label{eqn:recurseleaves}
\ell_a \; \stackrel{(i)}{=} \; \sum_{b \in \Children(a)} \frac{
  \ell_b}{|\Parents(b)|}, \quad \mbox{and} \quad m_a \;
\stackrel{(ii)}{=} \; 1 + \sum_{b \in \Children(a)} \frac{
  m_b}{|\Parents(b)|}.
\end{align}
In other words, the counts at each node are split evenly between its
parents, and so on. By construction, these counts satisfy the
following identities at the roots:
\begin{align}
  \label{eqn:identityleavesd}
\sum_{a \in \H_1} \ell_a \; \stackrel{(i)}{=} \; L, \quad \mbox{and}
\quad \sum_{a \in \H_1} m_a \stackrel{(ii)}{=} N.
\end{align}
The above calculation is analogous to a water-filling
procedure~\citep{lynch2014control,meijer2015multiple1}: imagine
pouring a unit of water into all the leaf nodes, and then turning the
graph upside down so that the leaves are at the top and roots at the
bottom. Then, water will flow due to gravity from leaves to roots
according to the dynamics~\eqref{eqn:recurseleaves}(i), whereas
equation~\eqref{eqn:identityleavesd}(i) corresponds to the
conservation of water at the roots.


\subsection{Generalized step-up procedures}

For the moment, suppose that we are only testing a batch of $K$
hypotheses. A threshold function for $P_i$ is mapping of the form $\alpha_i :
\{1,\dots,K\} \mapsto [0,1]$. Each function $\alpha_i$ is also implicitly a function of 
the constant target FDR level $\alpha$, but we leave this dependence implicit.
 A generalized step-up
procedure associated with a sequence of threshold functions
$\{\alpha_i(r)\}_{i=1}^K$ works as follows. It first calculates the
number of rejections as
\begin{align}
\label{eqn:stepupproc}
R & \defn \arg \max_{r = 1, \ldots K} \Big [ \sum_{i=1}^K \One{P_{i}
  \leq \alpha_i(r)} \geq r \Big],
\end{align}
and then rejects all hypotheses $i$ such that $P_{i} \leq
\alpha_i(R)$. For example, the Benjamini-Hochberg procedure is
recovered by using $\alpha_i(r) = \alpha r / K$ for all $i$. \DAGGER uses a generalized step-up procedure to test the
hypotheses within $\H_d$ for each $d=1,\ldots,D$, as described below.


\subsection{The \DAGGER algorithm}

Under independence or positive dependence,
  \DAGGER uses the generalized step-up procedure \eqref{eqn:stepupproc} with the functions
\begin{subequations}
\begin{align}
  \label{eqn:stepuplevels1}
\alpha_{d,i}(r) = \One{ \bigcap\limits_{j \in \Parents(i)} H_{d,j} \in
  \Rejections_{1:d-1} } \alpha \frac{\ell_i}{L} \frac{m_i + r
  + R_{1:d-1} - 1}{m_i}.
\end{align}
If we wish to protect against
arbitrary dependence, then we
choose reshaping functions $\beta_1,\dots,\beta_D \in \beta(\cT)$, and
 the reshaped \DAGGER algorithm uses procedure \eqref{eqn:stepupproc} with
\begin{align}
  \label{eqn:stepuplevels2}
\alpha_{d,i}(r) = \One{ \bigcap\limits_{j \in \Parents(i)} H_{d,j} \in
  \Rejections_{1:d-1} } \alpha \frac{\ell_i}{L} \frac{\beta_d(m_i + r
  + R_{1:d-1} - 1)}{m_i}.
\end{align}
\end{subequations}

Specifically, we run the generalized step-up procedure \eqref{eqn:stepupproc} on $\H_d$ for each $d \in \{1,
\ldots D \}$ with threshold functions
$\{\alpha_{d,i}(r)\}_{i=1}^{|\H_d|}$, meaning that we initialize
$\Rejections_{1:0} := \emptyset, R_{1:0} := 0$ and update
$\Rejections_{1:d} = \Rejections_d \cup \Rejections_{1:d-1}$ and
$R_{1:d} = R_d + R_{1:d-1}$, where
\begin{subequations}\label{eqn:rejections}
\begin{align}
R_d &= \max \left[ 1 \leq r \leq |\H_d| : \sum_{i=1}^{|\H_d|} \One{P_{i}
  \leq \alpha_{d,i}(r)} \geq r \right],\\ \Rejections_d &= \{ i \in \H_d :
P_i \leq \alpha_{d,i}(R_d) \}.
\end{align}
\end{subequations}
We refer to this top-down procedure as greedily evolving
  rejections on DAGs, or \DAGGER for short. An
  illustrative example of how \DAGGER works on a simple DAG
  is provided in the supplement. The exact choice of reshaping
function $\beta_d$ does not affect FDR control, but does affect the
power.  In proposing reshaping functions $\beta_d$, it is reasonable
to restrict to measures $\tau$ that put mass only on the values
that its argument could possibly take. For example, in order to mimic
the Benjamini-Yekutieli procedure, the reshaping function $\beta_d$ would assign mass proportional to
$1/k$ to each of the real numbers $k \in \{m_i + d - 1, m_i + d,
\dots, m_i + |\H_{1:d}| - 1\}$. This assignment occurs because
whenever there are any rejections at level $d$, we must have $d \leq r
+ R_{1:d-1} \leq |\H_{1:d}|$.  Special cases of \DAGGER include:
\begin{itemize}
\item First, suppose that the DAG is trivial, with $N$ nodes and no
  edges.  Then all hypotheses are leaves, and \DAGGER
  reduces to the Benjamini-Hochberg procedure under independence or
  positive dependence, and to the Benjamini-Yekutieli procedure under
  arbitrary dependence. 
%
\item Second, suppose that the DAG is a line graph.  On this DAG, the
  hypotheses are fully ordered, and our procedure reduces to the fixed
  sequential testing procedure of~\citet{lynch2016control}.
\item Third, consider a DAG in which each node has at most one parent,
  such as a tree (or a forest). In such a DAG, our procedure reduces to the
  hierarchical testing procedure of~\citet{lynch2016procedures}.
\end{itemize}

\noindent The following theorem lists some sufficient conditions under which
\DAGGER controls FDR. Below, we use $\fdr(\H_{1:d})$ to denote the
$\fdr$ achieved after the first $d$ rounds, so that $\fdr =
\fdr(\H_{1:D})$.

\begin{theorem}\label{thm:main1}
The \DAGGER algorithm (Eq.~\ref{eqn:stepupproc}, \ref{eqn:stepuplevels1}, \ref{eqn:rejections}) and the reshaped \DAGGER procedure (Eq.~\ref{eqn:stepupproc}, \ref{eqn:stepuplevels2}, \ref{eqn:rejections}) have the following guarantees:
\begin{itemize}
  \addtolength{\itemindent}{0.6cm}%
\item[(a)] (All DAGs) If all the $p$-values are either independent or
  positively dependent, then \DAGGER 
  guarantees that $\fdr(\H_{1:d}) \leq \alpha$ for all $d$.
\item[(b)] (All DAGs) If the $p$-values are arbitrarily dependent, then
  reshaped \DAGGER guarantees that $\fdr(\H_{1:d}) \leq \alpha$ for all $d$.
\item[(c)] (Intersection DAGs) If the $p$-values at the leaves are
  positively dependent, and other $p$-values are formed using Simes'
  procedure (either on the leaves of the node's sub-DAG, or directly on the node's children), then \DAGGER ensures that $\fdr(\H_{1:d}) \leq \alpha$ for all $d$.
\item[(d)] (Intersection DAGs) If the $p$-values at the leaves are
  independent or arbitrarily dependent, and all other $p$-values are formed using any valid
  global null test on the leaves (such as Fisher's, Rosenthal's, R\"uger's or R\"uschendorf's
  methods), then reshaped \DAGGER
  guarantees that $\fdr(\H_{1:d}) \leq \alpha$ for all $d$.
  \end{itemize}
\end{theorem}

\noindent The proof of this theorem, which is based on a bottom-up
inductive argument, is provided in the supplementary material. 
The proof strategy is
similar to that of the tree-FDR algorithm of
\cite{lynch2016procedures} that our work generalizes, but our proof
exploits recent work on super-uniformity lemmas developed both in the offline
context by \cite{blanchard2008two} and \cite{ramdas2017unified}, and in the online context by \cite{JM16} and \cite{RYWJ17}.

The \DAGGER algorithm and proof combine the structure of the DAG with ideas from
batch FDR procedures as well as online FDR procedures. Naturally, the
use of generalized step-up procedures is reminiscent of batch FDR
procedures such as the BH method. Further, if we make more rejections
early on, then we are allowed to test later hypotheses
at more lenient thresholds, because $\alpha_{d,i}(r)$ increases with
$R_{1:d-1}$, reminiscent of online FDR procedures that earn alpha-wealth
upon making rejections, to spend on later tests.  
Note that for
trees, the FWER procedure by \cite{meinshausen2008hierarchical} tests
each node whose parents were rejected at a level $\alpha
\frac{\ell_i}{L}$, and hence the last term in the expression for
$\alpha_{d,i}(r)$ intuitively captures the more lenient thresholds
that \DAGGER enjoys since it is always larger than one.



\section{Simulations}\label{sec:sims}
Here, we compare \DAGGER
with a variety of existing algorithms, including
\begin{itemize}
\item
the \texttt{Focus-level} method of \citet{goeman2008multiple};
\item
two generalizations of the procedure of \cite{meinshausen2008hierarchical} to DAGs due to
\citet{meijer2015multiple1}, called \texttt{MG-b1} and \texttt{MG-b2};
\item
a \SHOLM procedure by \citet{meijer2015multiple2};
 \item
the two top-down procedures called \SCRDAG and \BHDAG by
\cite{lynch2014control};
\item the \LORD algorithm of \citet{JM16,RYWJ17}.
\end{itemize}
Note that the first four algorithms control FWER, while the last three
algorithms control FDR in the case of independent
$p$-values.The power and FDR are calculated by averaging the true discovery proportion
  \mbox{$|\Rejections \cap \mathcal{H}_1|/|\mathcal{H}_1|$} and the
  false discovery proportion \mbox{$|\Rejections \cap
    \mathcal{H}_0|/|\Rejections|$} over 100 independent runs. Code
for reproducing key empirical results is available
 online at \texttt{https://github.com/Jianbo-Lab/DAGGER/}.

We remind the reader that
 \DAGGER is the only algorithm that works under
  motivations (M1) and (M2), being both sequential and handling
  arbitrary dependence. Hence, in order to have at least a few
  algorithms to compare to, we only consider independent $p$-values
  and only motivation (M2), meaning that all $p$-values are made
  available offline so that algorithms taking multiple passes over the
  $p$-values are allowed. Furthermore, it should be noted that online
  FDR algorithms like \texttt{LORD}, an instance of a monotone generalized
    alpha-investing rule for which FDR control has been proved in past
    work \citep{JM16,RYWJ17,aharoni2014generalized,FS08}, were originally
  designed for testing fully ordered hypotheses in an online manner, and none of the
  existing online FDR papers consider applications to DAGs. Indeed,
   we apply \texttt{LORD} here by converting the
  DAG into a full ordering in a somewhat ad-hoc manner,
  first testing the roots (in some arbitrary order), then testing
  nodes whose parents were rejected (in some arbitrary order), and so
  on.  


\subsection{Results and discussion of experiments}

Before providing the details of all the simulations and the real data
problem, we first summarize some high-level takeaway messages from the
various experiments.

 Results under motivation (M2):  \DAGGER is very
  computationally efficient, requiring only one simple sort operation
  per depth. \LORD is slightly faster, since it does not even sort
  $p$-values. However, the other FDR algorithms, in particular
  \texttt{SCR-DAG} and \texttt{BH-DAG}, are slower by several orders
  of magnitude, running several hundred times slower than
  \DAGGERSHORT. This is because each iteration of these algorithms makes a
  pass over the entire DAG, and the number of such iterations equals
  the number of ultimately unrejected hypotheses. For the FWER
  algorithms, the \texttt{Focus-level} algorithm is prohibitively slow
  to repeat hundreds of times for each parameter setting, so we
  compare to it only in the real-data example in \secref{real}.  The
  other FWER algorithms are as fast as \DAGGERSHORT. The fastest
  algorithms are \SHOLM which typically has the lowest power, and the
  BH procedure which has the highest power since it 
  ignores structural constraints and rejects hypotheses anywhere on
  the DAG including nodes whose parents are not rejected.

Across simulations, the FDR methods usually have greater
  power than the FWER methods. While it might be expected that
  Benjamini-Hochberg would have more power than \DAGGER because it is
  unconstrained, it turns out that this is not necessarily the case,
  and we provide a counterexample in this section. In the
  supplementary material, we provide intuition about how the power of
  \DAGGER changes with the DAG shape.

In summary, we find that \DAGGER performs well in
  simulations both in terms of time and in terms of power. It has the
  advantage of being applicable under motivation (M1) and being able
  to handle the kinds of dependence listed as (D1)-(D6) in
  \secref{alg1}.

 \LORD versus \DAGGERSHORT~under motivation (M1): Even though
\LORD was adapted to the setting of partially ordered hypotheses in a
somewhat ad-hoc manner (arbitrarily converting a partial ordering into
a full ordering), our simulations show that it performs reasonably
well in practice. Since \LORD and \DAGGER are the only two algorithms
(of which we are aware) that provide FDR control under motivation
(M1), it is instructive to compare their strengths and weaknesses.

Both algorithms have the property that past rejections allow future
hypotheses to be tested at larger levels.  The main advantage of \LORD
is that it does not waste any alpha-wealth on testing hypotheses whose
parents were not rejected, but it faces three significant
disadvantages compared to \DAGGERSHORT. Firstly, \DAGGER tests each
node at a different threshold that depends on the number of its
descendants. As an example, if there are two nodes at the same depth,
one being a leaf, and the other having thousands of descendants, then
\DAGGER will assign a larger test threshold to the latter node because
of the number of future tests that depend on it, while \LORD cannot
adjust threshold levels for each test to take advantage of the DAG
structure.

Secondly, \DAGGER can test all the levels at a particular depth as a
single batch, possibly gaining power by treating the $p$-values
together using the generalized step-up procedure, as opposed to
\texttt{LORD}, which cannot take other $p$-values into account when
making each of its decisions. Lastly, unlike \texttt{LORD}, \DAGGER does not
need to conservatively reshape thresholds when dealing with positive
dependence among $p$-values on general DAGs, and for intersection
DAGs, it can handle positive dependence between base $p$-values
when Simes' $p$-values are used at other nodes.

Unfortunately, not much can be said in general to compare the power of the
two algorithms, since it is affected by the choice of graph structure,
positions of the non-nulls and the density of the non-null $p$-values. In all our
experiments, \DAGGER was more powerful than \LORD.


\subsection{Comparing various methods for FWER and FDR control}

 Graph structure:  We follow the simulation protocol of
\citet{meijer2015multiple1} and choose the Gene Ontology and one of
its subgraphs as the underlying graph structure.  The Gene Ontology is
a DAG where each node represents a gene ontology term and the
relationships between the terms are represented by edges between the
nodes. The child nodes represent more specialized terms than parent
nodes. The gene ontology graph has $4952$ nodes in total. We also used
the subgraph of the gene ontology with cell proliferation as its top
node with $494$ nodes.

 Distribution of null and alternative hypotheses:   For each
replication of our experiments, we distributed the null hypotheses and
the alternative hypotheses randomly. The leaves are randomly chosen
with probability $\pi_0^L$ to be true null hypotheses. The rest of the
hypotheses are assigned to be true nulls if and only if all of their
children are true nulls. We generate $p$-values in the following two
different manners:
\begin{itemize}
\item (Independent $p$-values.)
In the first setting, each $p$-value is independently generated as
\begin{align}
  \label{$p$-value-model}
X\sim \mu + \mathcal{N}(0,1); \text{ $p$-value} = 1-\Phi(X),
\end{align}
where $\Phi$ is the standard Gaussian CDF, with $\mu=0$ for nulls and
$\mu>0$ for alternatives. Larger values of $\mu$ indicate stronger
signals. We decrease the signal of alternatives linearly with increasing depth. Concretely, we set $\mu=1$ at the nodes of the largest depth and increase the signal by $0.3$ as depth decreases. 

\item (Simes' $p$-values.)
In the second setting, $p$-values in leaves are independently generated from the previous model, with $\mu=2$ at alternatives. Every other node's $p$-value is calculated recursively as the Simes' $p$-value of the node's children.
\end{itemize}

 Parameters: For each setting, we increase the proportion of
nulls on leaves $\pi_0^L$ over the entire DAG from $0.15$ to $0.95$ by
$0.05$ and plot the power for each algorithm. We repeat each
experiment $100$ times. We set $\alpha=0.2$, which refers to the
target FWER for \texttt{MG-b1}, \texttt{MG-b2} and \SHOLM and the
target FDR for \texttt{BH}, \texttt{SCR-DAG}, \BHDAG and
\DAGGERSHORT. As \texttt{SCR-DAG} and \BHDAG do not work in the case
of arbitrary dependence, we run these algorithms only in the
independent setting. \LORD requires specification of an infinite
sequence of constants, which we choose to be the same as those used in
prior work for mixtures of Gaussians~\citep{JM16,RYWJ17}, since the
first paper justifies the heuristic optimality of this choice by
deriving a lower bound on power in unstructured settings.

\begin{figure}[h!] 
\centering
    \includegraphics[width=.48\linewidth]{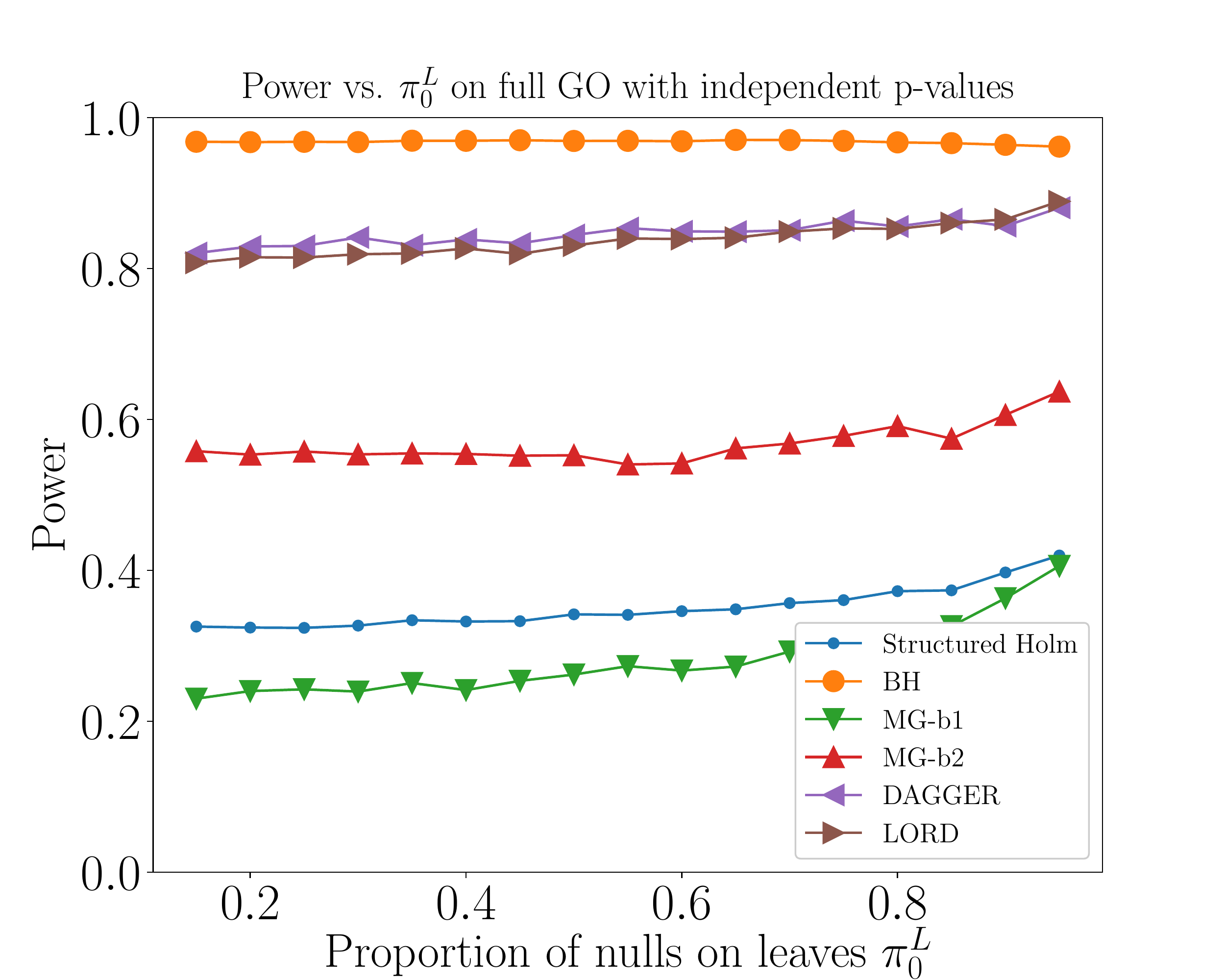}
    \includegraphics[width=.48\linewidth]{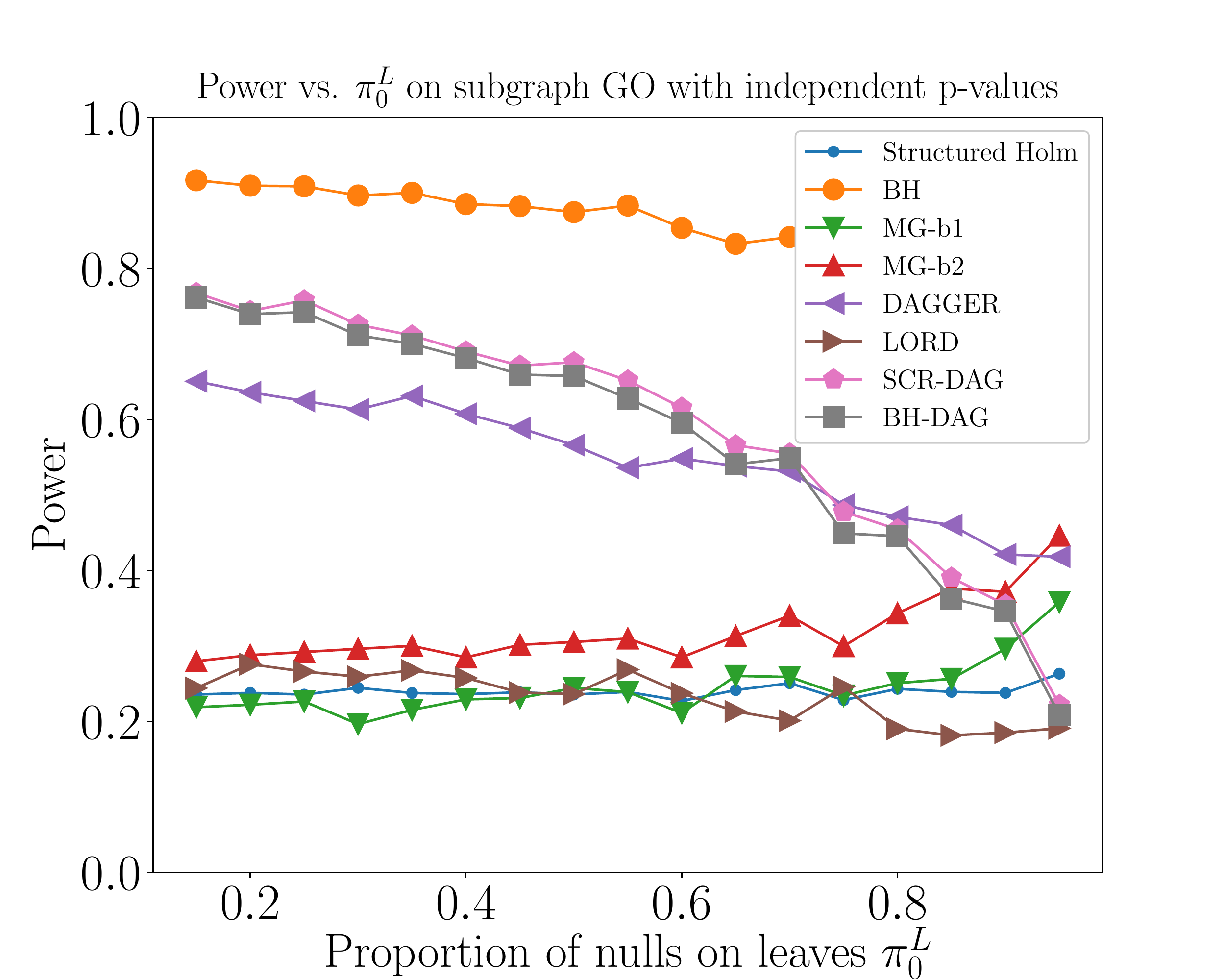}
    \includegraphics[width=.48\linewidth]{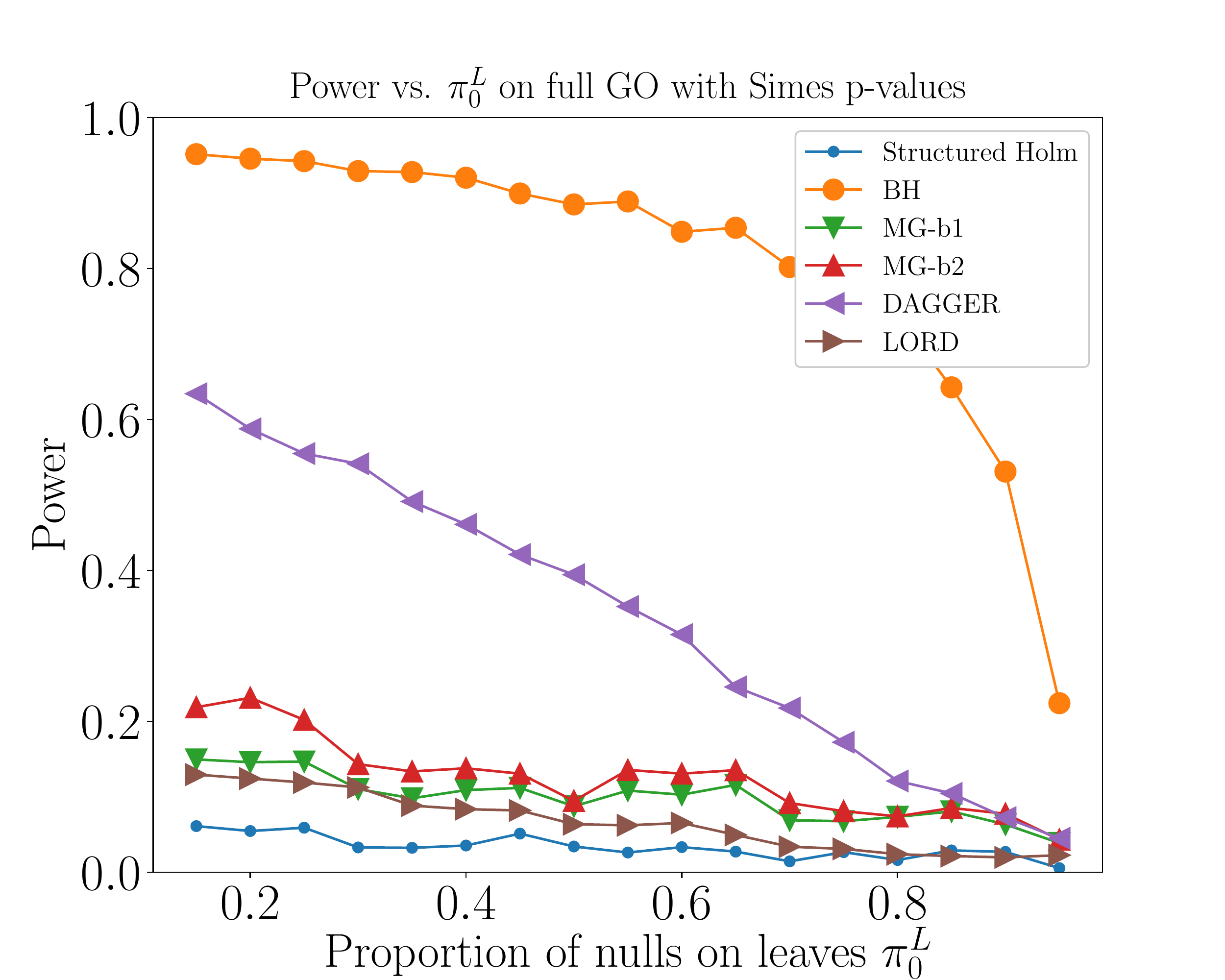}
    \includegraphics[width=.48\linewidth]{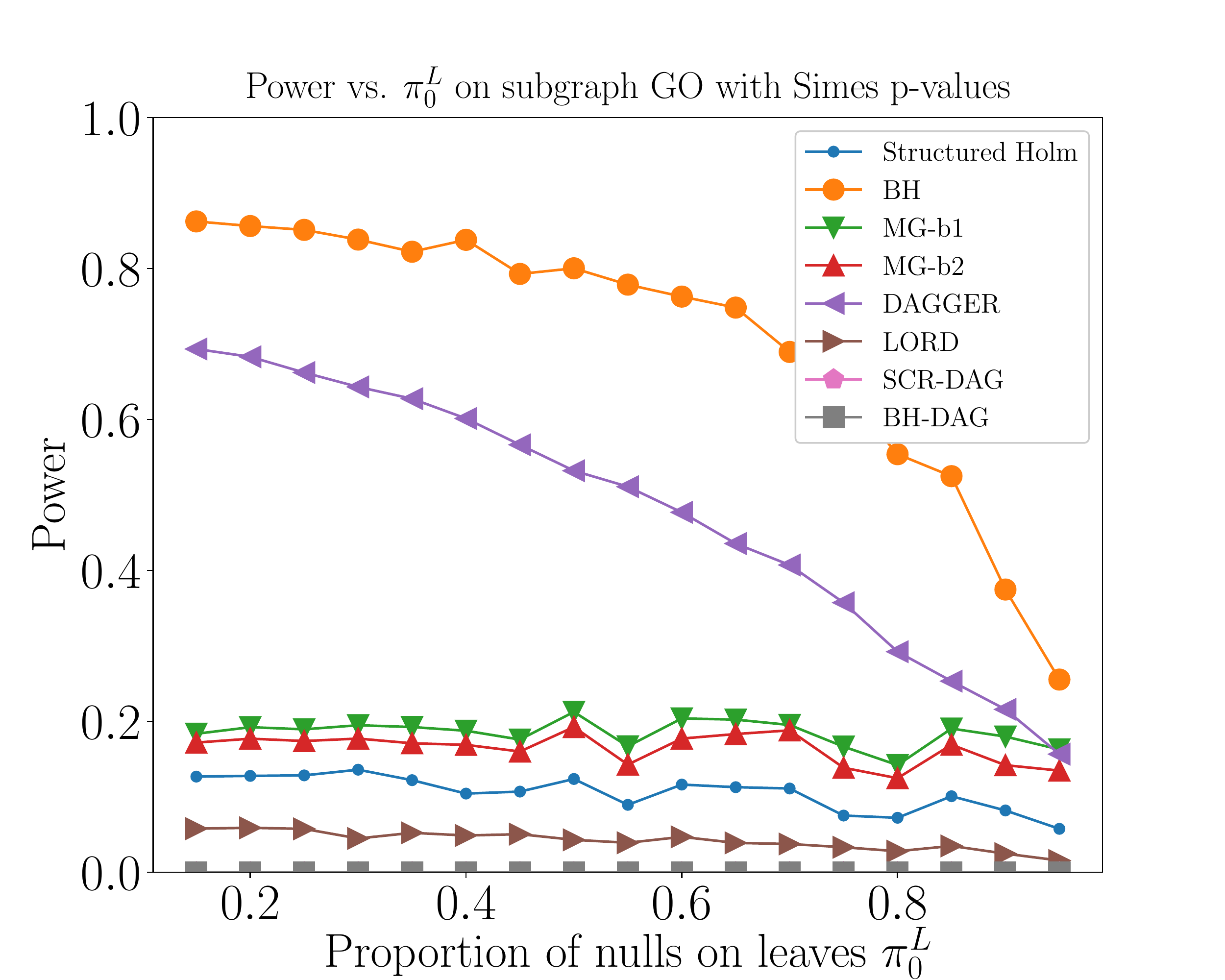} 
    \caption{Power of various algorithms on the entire gene ontology
      graph (left column) and the subgraph rooted at cell
      proliferation (right column), under the setting of independent
      $p$-values (top row) and Simes' $p$-values (bottom row). }
\label{fig:independent+Simes}
\end{figure}

\begin{figure}[h!] 
\centering
    \includegraphics[width=.49\linewidth]{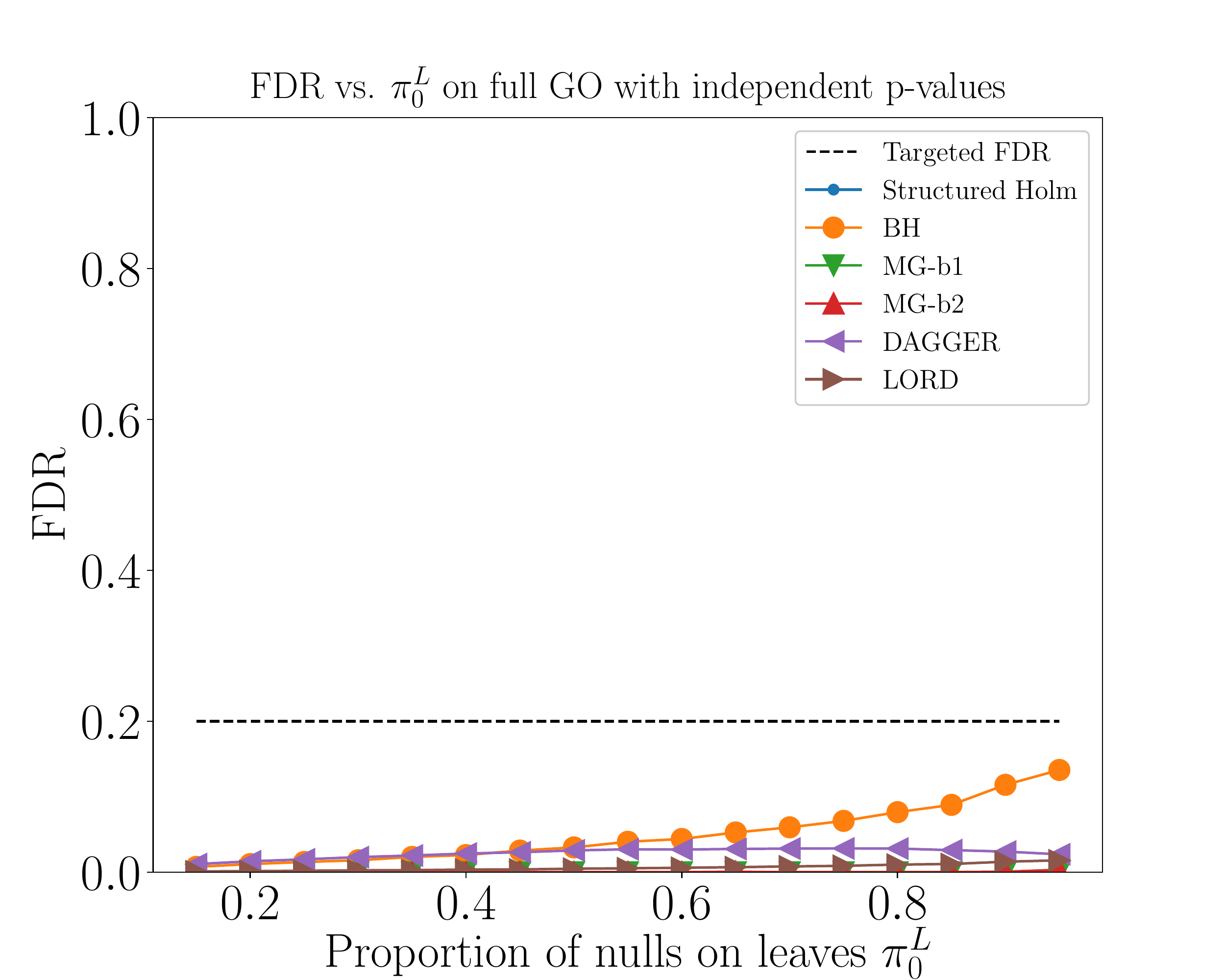}
    \includegraphics[width=.49\linewidth]{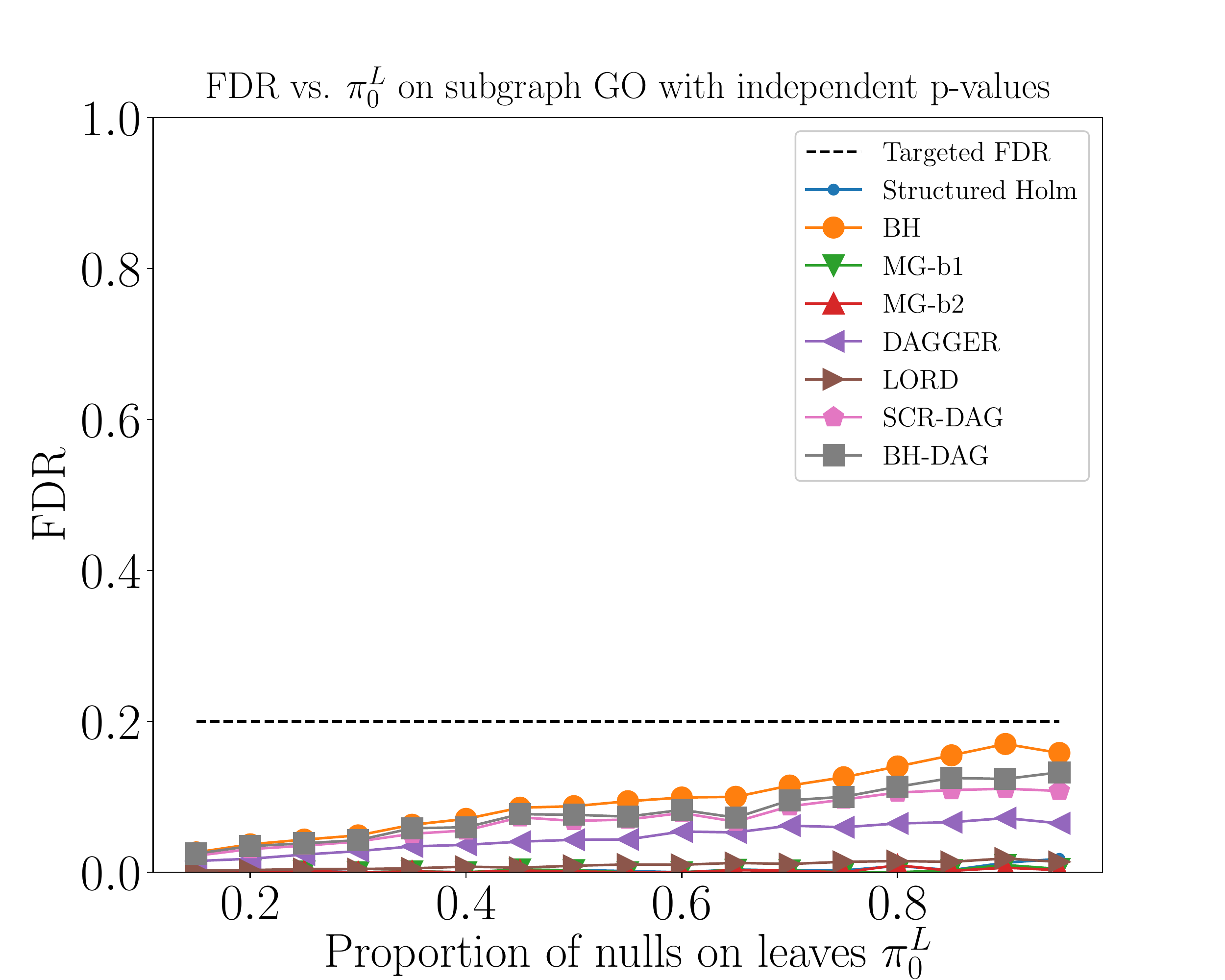}
    \includegraphics[width=.49\linewidth]{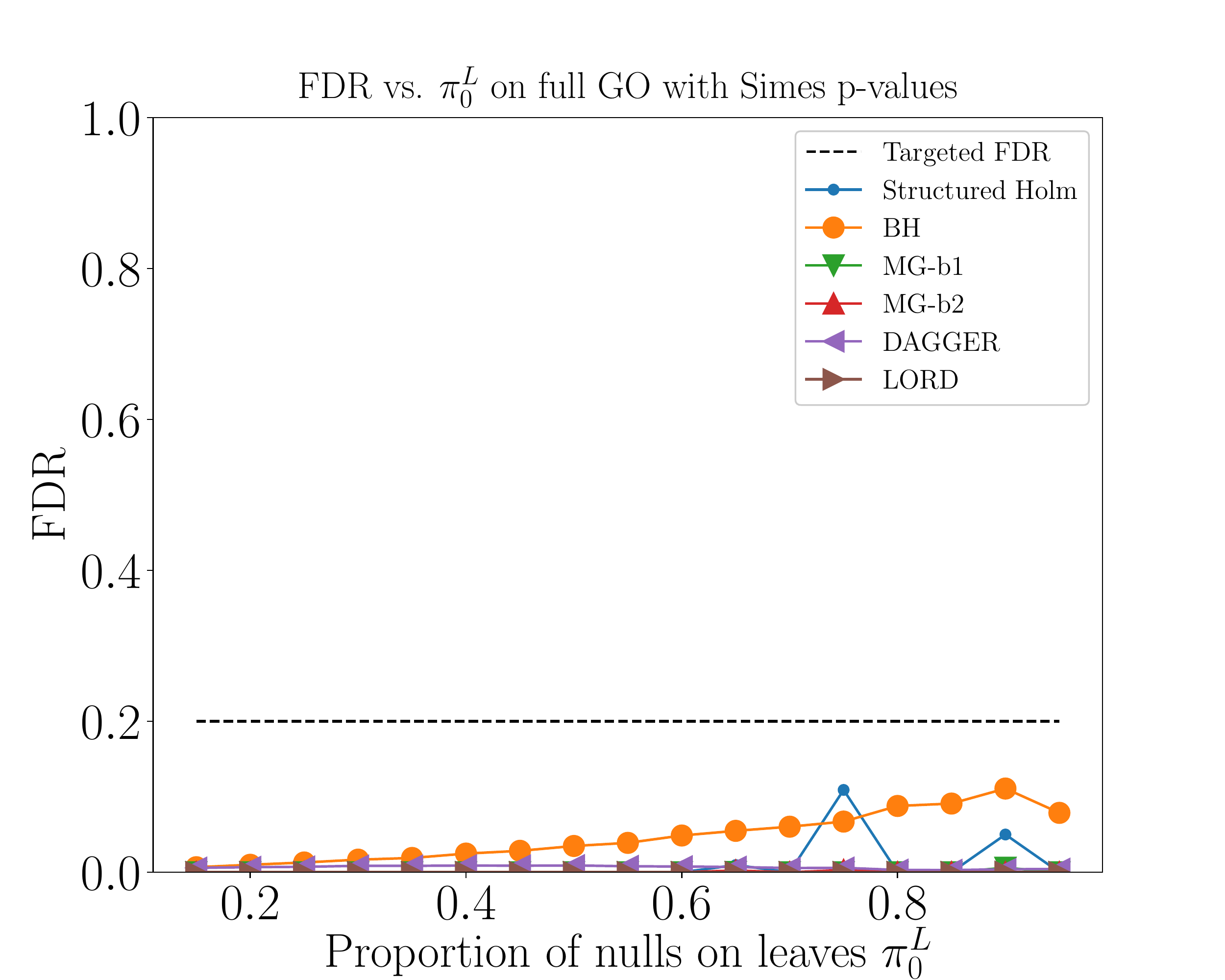}
    \includegraphics[width=.49\linewidth]{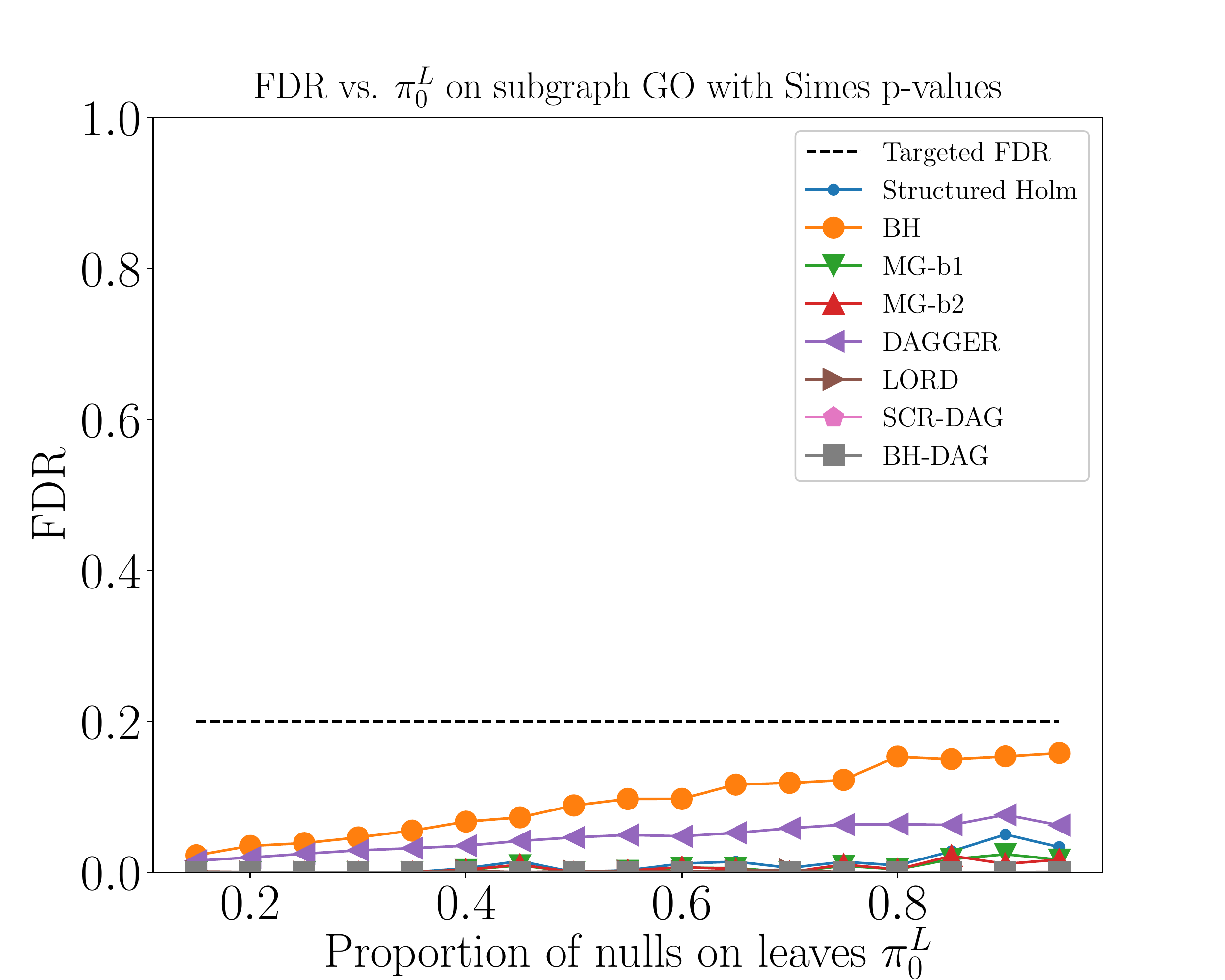} 
    \caption{Plots of the achieved FDR versus proportion of nulls on leaves
      for various algorithms for simulations on the entire gene ontology
      graph (left column) and the subgraph rooted at cell proliferation
      (right column), under the setting of independent $p$-values (top row)
      and Simes' $p$-values (bottom row). }
\label{fig:fdr}
\end{figure}

 Results: 
From \figref{independent+Simes}, the Benjamini-Hochberg method, which controls FDR and does not take logical constraints into consideration, has the largest power as expected. Our algorithm is more powerful than \SCRDAG and \BHDAG in the case where there are a large number of nulls, which is a relatively more common and practical setting, and the latter algorithms are also non-sequential, do not work with arbitrary dependence and have a much bigger time complexity, as discussed below. \LORD is less powerful than \texttt{SCR-DAG}, \BHDAG  and \DAGGER on the subgraph, but achieves larger power on the full gene ontology graph. 
Finally, all the algorithms controlling FDR have larger power than the algorithms that control FWER when there are relatively large number of nulls. 
We refer the reader to Figure \ref{fig:fdr} for the achieved FDR of various algorithms.

\begin{figure}[h!] 
\centering
\includegraphics[width=0.48\linewidth]{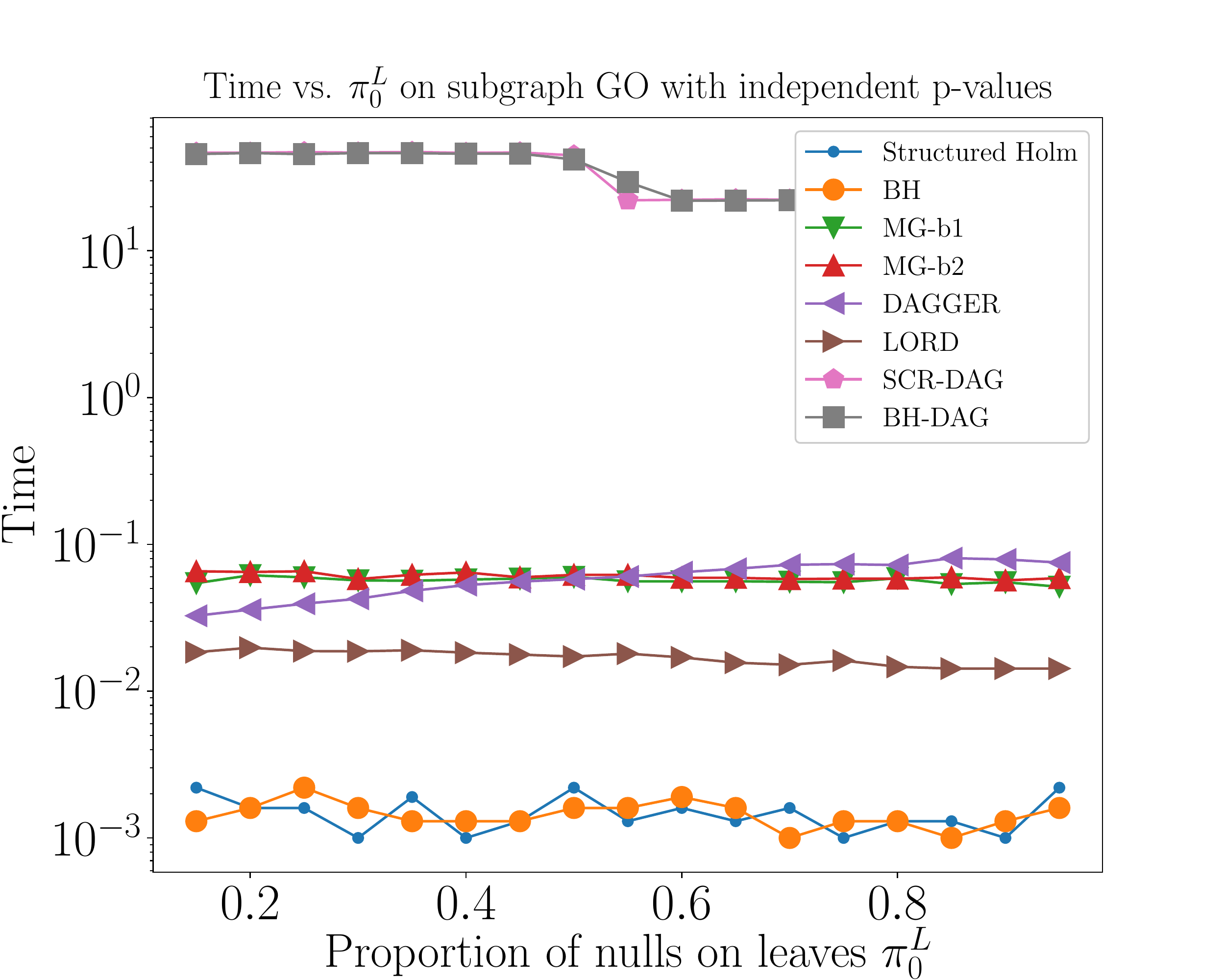}
  \caption{Clock time of algorithms on the subgraph of the gene
    ontology DAG for independent $p$-values. The simplicity of
    \DAGGERSHORT~makes it readily scalable to large graphs.}
\label{fig:complexity}
\end{figure}

 Time complexity: Figure \ref{fig:complexity} shows the
clock-time of various algorithms on a log scale. As the two top-down
procedures of Lynch are too time-consuming to run over the whole gene
ontology graph, we only report their results on subgraphs of the gene
ontology. In summary, \SHOLM and Benjamini-Hochberg are the fastest,
followed by \texttt{LORD}, \DAGGERSHORT, \texttt{MG-b1} and
\texttt{MG-b2}. Lynch's \SCRDAG and \BHDAG are the slowest, as is the
\texttt{Focus-level} algorithm tested in the real data subsection,
making these impractical for large DAGs.


\begin{figure}[h!] 
    \includegraphics[width=.49\linewidth]{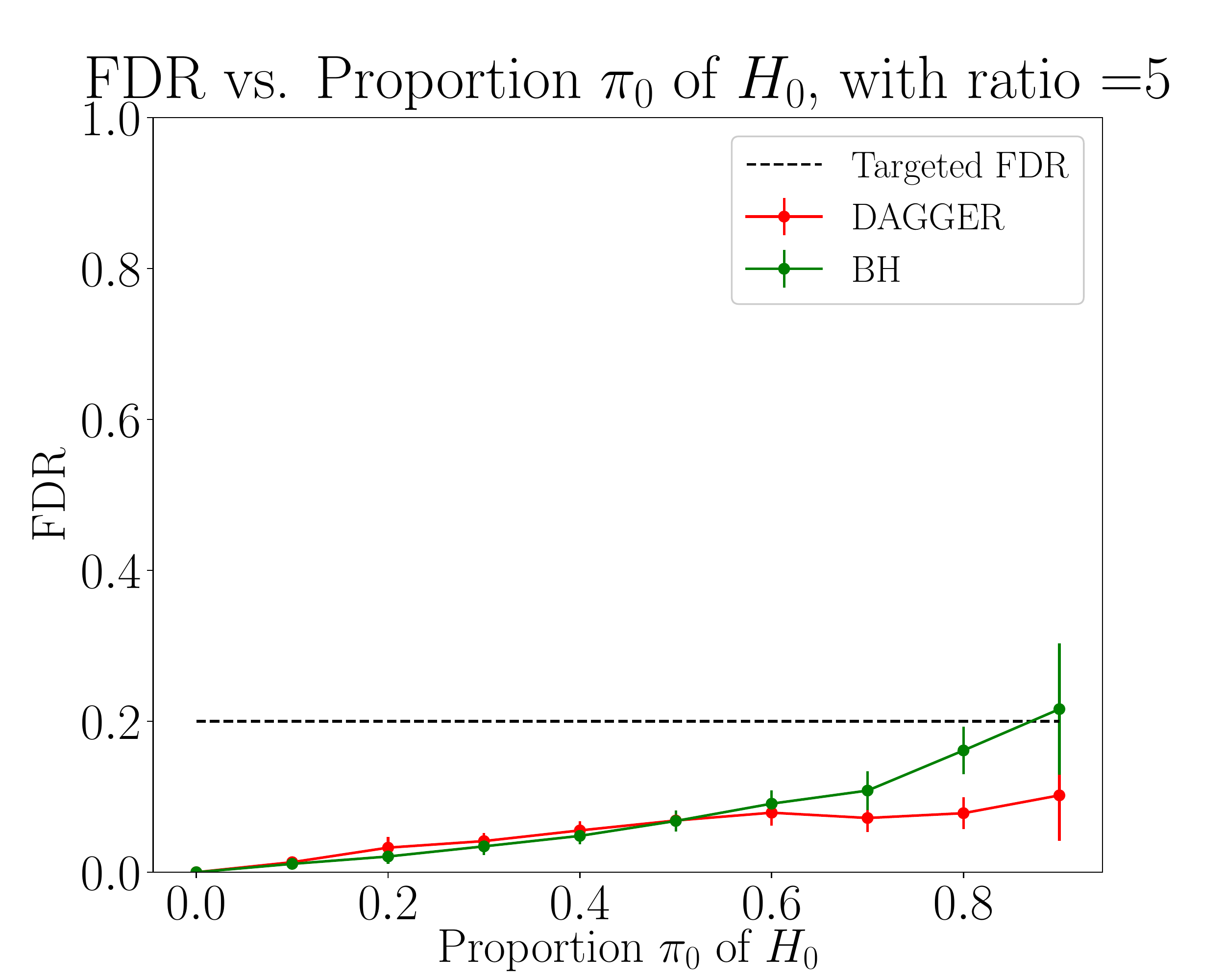}\quad
    \includegraphics[width=.49\linewidth]{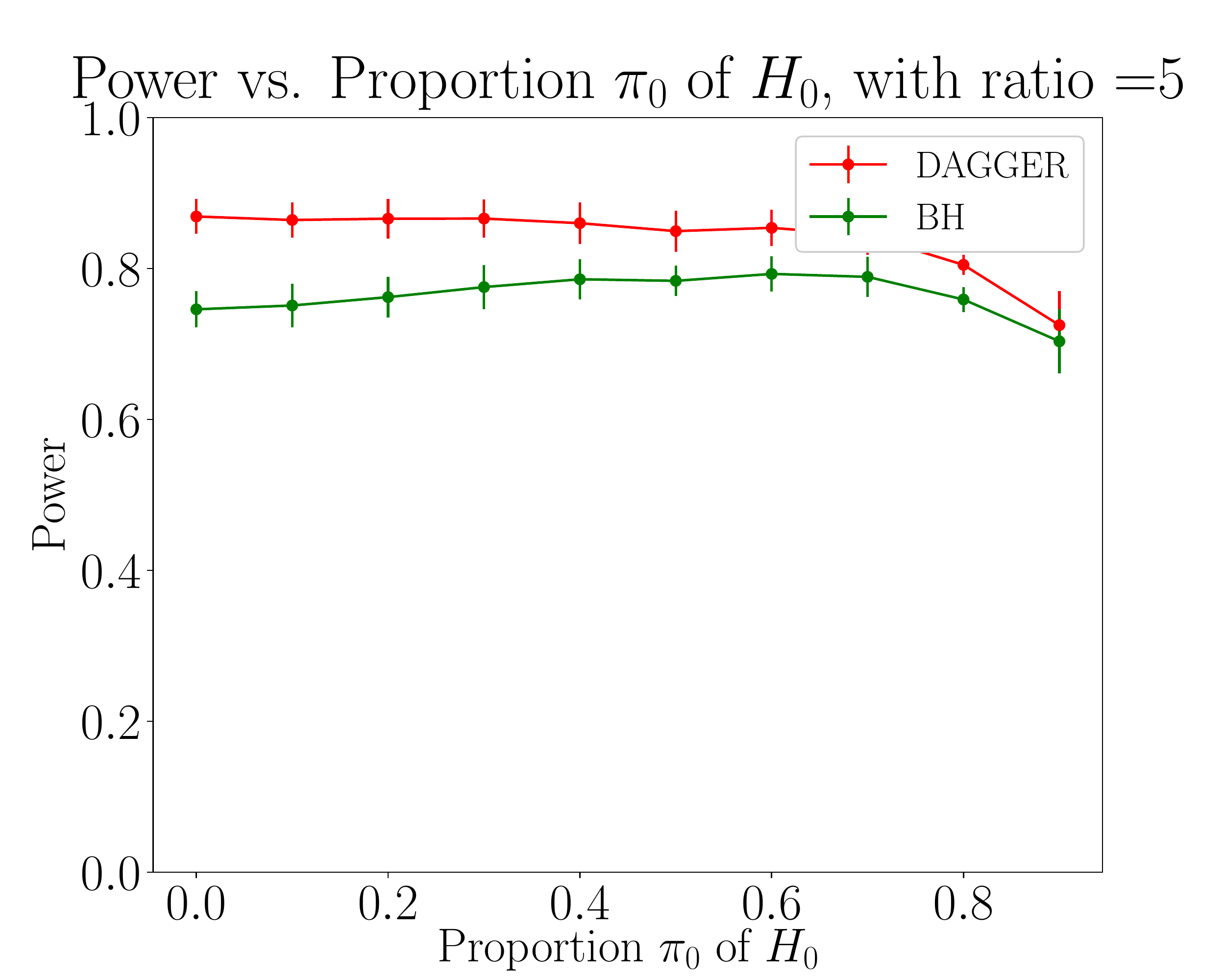} 
  \caption{Plots of the achieved FDR and power versus the proportion of nulls in the case where signals are stronger at the top layer, controlling the target FDR at $\alpha=0.2$, the number of nodes $n=200$ and with $100$ repetitions. } 
\label{fig:GSU-BH1}
\end{figure} 

\subsection{A case where \DAGGER is more powerful than BH}

In the remaining simulations in the paper and supplement, each
$p$-value $P$ is independently generated from the model
\begin{align}
  \label{$p$-value-model}
X\sim \mu + \mathcal{N}(0,1) \quad \mbox{and} \quad P = 1-\Phi(X),
\end{align}
where $\Phi$ is the standard Gaussian CDF, with $\mu=0$ for nulls and
$\mu>0$ for alternatives. Larger values of $\mu$ indicate stronger
signals. We fix the target FDR to be $\alpha=0.2$ for all experiments.

Unlike the Benjamini-Hochberg algorithm, the rejections made by
\DAGGER respect the graph structure, and so we might expect
Benjamini-Hochberg to always have higher power. Here we describe a
setting where \DAGGER actually achieves higher power than
Benjamini-Hochberg. It is when the non-nulls in the layers of smaller
depth have stronger signal strength than the non-nulls in the deeper
layers. Concretely, we generate a DAG of depth $2$ with each node in
the bottom layer randomly assigned to $2$ nodes in the top layer. The
signal strengths of non-nulls are $\mu=5$ and $\mu=1$ in the top and
the bottom layer respectively. We control the target FDR to be
$\alpha=0.2$, and assign $100$ nodes to the bottom layer and $100$
nodes to the top layer. Each experiment is repeated $100$ times.

Figure \ref{fig:GSU-BH1} shows how the achieved FDR and the power vary
with the proportion of nulls in the entire DAG. The error bars
indicate in-sample standard deviations. The achieved power of \DAGGER
is larger than the achieved power of the Benjamini-Hochberg procedure. We
observe that both Benjamini-Hochberg and \DAGGER achieve a much lower
FDR than the targeted FDR, opening up the possibility of adaptive
algorithms using null-proportion estimates
\citep{Storey04,ramdas2017unified} if independence is assumed.


\section{An application to the Gene Ontology DAG}\label{sec:real}

In this section, we present the results of an experiment comparing the
\DAGGER method with \texttt{Focus-level} methods
\citep{goeman2008multiple}, the reshaped Benjamini-Hochberg method by
\cite{BY01} that handles arbitrary dependence in unstructured
settings, and all the other algorithms used in the simulation section.

As done in past work, we use a subset of the Golub dataset
\citep{golub1999molecular,golub2016data} that contains $72$ tumor mRNA
samples in which the expression of $7129$ genes is
recorded. Specifically, the Golub data set is from the leukemia
microarray study, recording the gene expression of $47$ patients with
acute lymphoblastic leukemia and $25$ patients with acute myeloid
leukemia. We focus on the biological process ``cell cycle''
(GO:0007049) and its descendants in the Gene Ontology graph, a DAG of
$307$ nodes in total.

The gene ontology graph represents a partial order of the gene
ontology terms, and the set of genes annotated to a certain term
(node) is a subset of those annotated to its parent node
\citep{ashburner2000gene}. We test for differential biological process
activity between the two kinds of patients. Concretely, the null
hypothesis at each node is
\begin{align}
H_0: \text{ No gene in the node's gene set is associated with the type
  of diseases.}
\end{align}
Multiple testing procedures on the gene ontology graph are expected to
preserve the graph structure of Gene Ontology
\citep{goeman2008multiple}, meaning that a child node is rejected only
if all of its parents are rejected, which holds true for \DAGGERSHORT.

 Construction of $p$-values: Individual (raw) $p$-values on
each node are obtained by Global Ancova
\citep{doi:10.1093/bioinformatics/btm531}, which are calculated using
the GlobalAncova package in R \citep{hummel2016global}. The test
is carried out by comparison between the linear model containing all
covariates and the reduced model containing the covariates of
interest, which are the genes in the corresponding node here, via the
extra sum of squares principle. The test statistic involves residual
sums of squares from both the full model (FM) and the reduced model
(RM):
\begin{align}
F_{\text{GA}} = \text{Const.}\cdot\frac{\text{RSS}_{\text{RM}} -
  \text{RSS}_{\text{FM}}}{\text{RSS}_{\text{FM}}},
\end{align}
and the $p$-values are computed with a permutation-based approximation
\citep{doi:10.1093/bioinformatics/btm531}. Note that the $p$-values
obtained by are arbitrarily dependent, and we consequently adopt the
Benjamini-Yekutieli reshaping function in \texttt{DAGGER}  and \texttt{LORD},
which we call reshaped \DAGGER and reshaped \texttt{LORD}.

\begin{figure}[h!] 
\centering
\includegraphics[width=.8\linewidth]{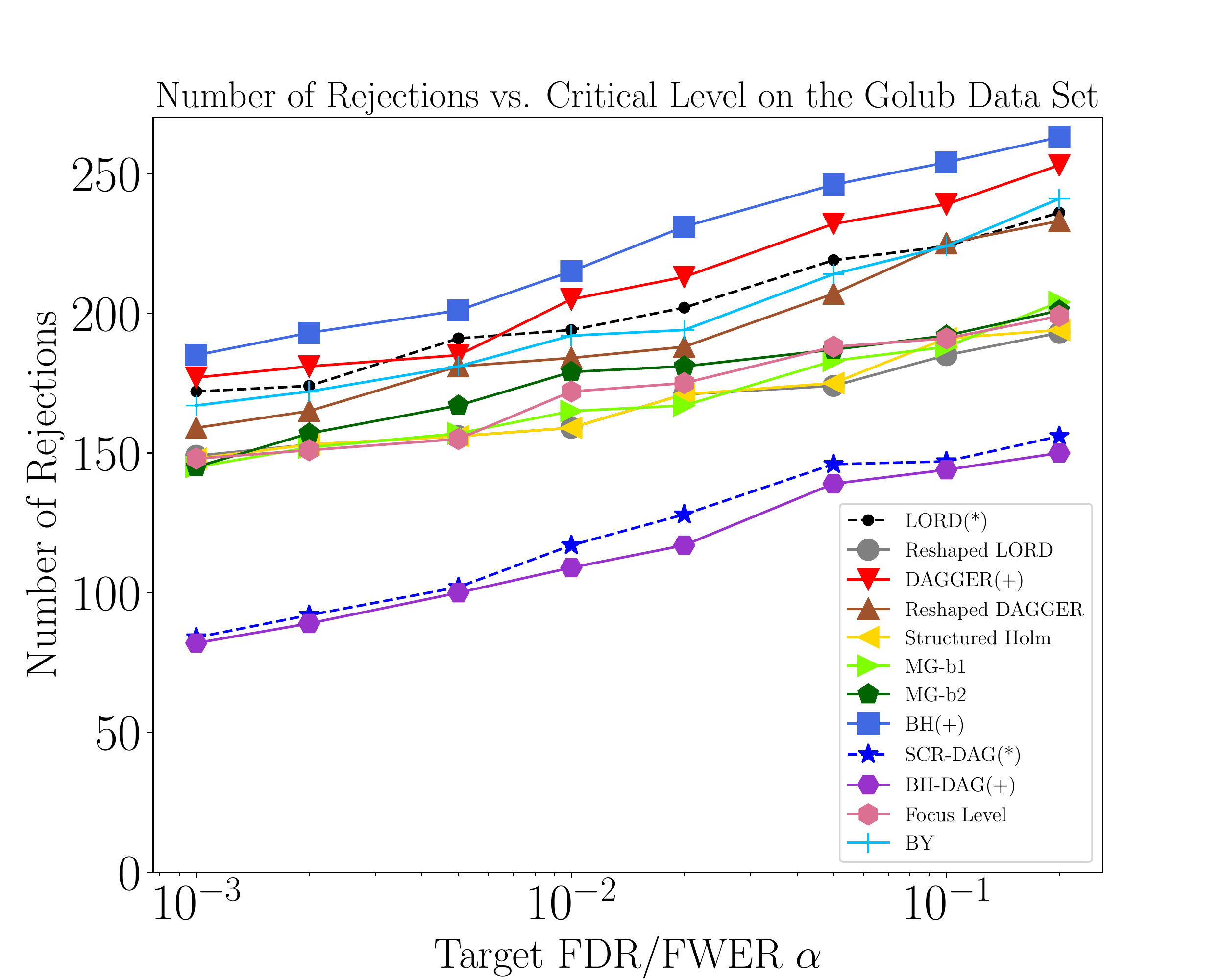}
  \caption{Plots of the number of rejections with various levels of
    $\alpha$: $\{0.001,0.002,0.005,0.01,0.02,0.05,0.1,0.2\}$, where
    $\alpha$ refers to the target FDR or FWER. The algorithms that
    work under an independence assumption (unrealistic for this
    dataset) are shown with a dotted line and followed by $(*)$ in the
    legend. The algorithms that work with the assumption of positive
    dependence (plausible, but not provable) are followed by $(+)$ in
    the legend. Other algorithms make no dependence assumptions.
    Among those methods that maintain the DAG structural
      constraints under arbitrary dependence, or even assuming
      positive dependence, \DAGGERSHORT~is the most powerful
      algorithm. }
\label{fig:GO_num_rejs}
\end{figure} 

\begin{figure}[h!]  
\centering \includegraphics[width=0.75\linewidth]{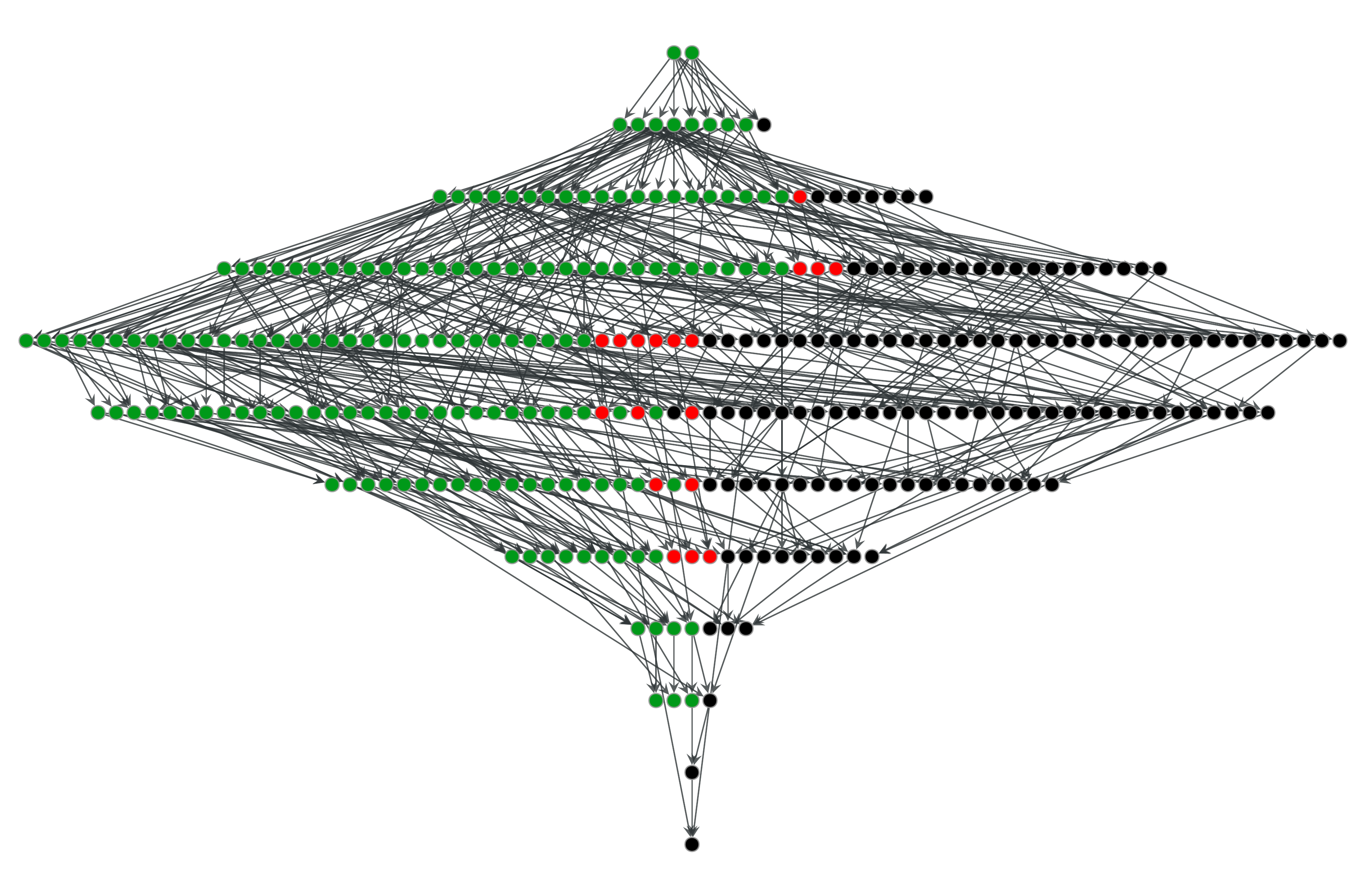}
  \caption{Green nodes (size 177) are rejections made by reshaped
    \DAGGERSHORT, while red nodes (size 18) are the additional
    rejections made by \DAGGER if positive dependence is assumed, both
    at a target FDR of $\alpha=0.001$.}
  \label{fig:GO}
\end{figure}  
 
 Compared algorithms:  \figref{GO_num_rejs} shows the number
of rejections at various critical levels $\alpha$, where $\alpha$
refers to the target FDR levels for \DAGGERSHORT, \texttt{LORD},
reshaped \texttt{DAGGER}, reshaped \texttt{LORD}, \texttt{SCR-DAG},
\texttt{BH-DAG}, the Benjamini-Hochberg and Benjamini-Yekutieli procedures, whereas $\alpha$ refers
to the target FWERs for \texttt{structured-Holm}, \texttt{MG-b1},
\texttt{MG-b2} and \texttt{Focus-level}. These $p$-values do not have the
assumption of independence or positive dependence, but we include
algorithms with these assumptions for comparison. Algorithms that only
control FDR under an independence assumption are shown with a dotted
line and are annotated by $(*)$ in the legend; these are \LORD and \texttt{SCR-DAG}.
Algorithms that control FDR under an assumption of positive dependence are 
annotated by a $(+)$ symbol in the legend; these are \DAGGERSHORT, \texttt{BH-DAG} and
\texttt{BH}. All other methods control FDR/FWER without dependence
assumptions.

 Time:   We apply the \texttt{Focus-level} method with the
default focus level in the GlobalAncova package
\citep{hummel2016global}, which takes about four minutes to run. We
note that the \texttt{Short-Focus-Level} procedure, recently proposed
as a more computationally efficient algorithm to control FWER on the
gene ontology graph by \cite{saunders2014shortcut}, takes about one
second. The rest of the algorithms are much faster in terms of clock
time and in particular,  \DAGGER takes about $0.02$ seconds,
indicating scalability to much larger graphs.

 Power:  Among the algorithms that control FDR under no
dependence assumptions while resepcting the strong hierarchical
principle, namely reshaped \texttt{DAGGER}, reshaped \LORD and all the
FWER algorithms, we see that reshaped \DAGGER is the most
powerful. Similarly, among algorithms that control FDR under the
assumption of positive dependence and also respect the structural constraints, we
find that \DAGGER again consistently yields the highest power.

It is worth noting that \DAGGER performs only slightly worse than Benjamini-Hochberg, while the reshaped \DAGGER performs almost as well as Benjamini-Yekutieli.

\figref{GO} shows the rejections made by \DAGGER and by reshaped
\DAGGER at $\alpha = 0.001$ respectively.  By construction, the set of
rejections by \DAGGER always contains all of the rejections made by
reshaped \DAGGERSHORT, but interestingly, in this example, one does
not lose much power by removing the assumption of positive dependence.


\section{Summary}
\label{sec:conclusion}

We have presented a sequential algorithm that exploits prior structural
knowledge in the form of a general DAG on hypotheses to selectively
carries out tests as needed, ultimately rejecting a sub-DAG that
satisfies the strong hierarchy principle.  We have shown that for this
algorithm FDR is controlled at a prespecified level under settings of independence, 
positive dependence or arbitrary dependence among the $p$-values. We have demonstrated the
utility of our procedure in simulations and on real data.

There are many promising theoretical as well as practical directions for future work, including designing procedures for other error metrics as well as other types of logical constraints. 
For example, it may also be of importance to consider logical
constraints beyond the strong hierarchical principle, by changing (C2) to other natural options,
for example by changing the ``all'' to ``any,'' or to other
pre-defined Boolean operators.
Also note that we have assumed that the DAG is finite and known in advance to the scientist. 
An important direction for future work under motivation (M1) is to design sequentially rejective
algorithms, that work on DAGs whose size and structure is not known in
advance, but are revealed as the algorithm proceeds down the DAG. 

There are other plausible applications that fall under the umbrella of
motivation (M2) beyond the gene ontology example that we considered. For example, in factorial experiments, the variable selection problem can be posed in terms of testing on a DAG: here one would consider testing for joint effects with only pairs or triplets of variables in which at least one of the marginal effects was already deemed significant being involved.  

\subsection*{Acknowledgments} 
We thank Jelle Goeman for helping reproduce results in his papers, and
 for constructive comments on an initial draft. We thank
Lihua Lei for sharing code for \texttt{SCR-DAG} and \texttt{BH-DAG}. 
This work was improved by comments from
  audiences of the MCP 2017 conference and Biostatistics
  seminars at Berkeley and Stanford.
  We also thank the reviewers for their comments
  that helped improve the paper.

\subsection*{Supplementary material}
Supplementary material includes an illustrative example of \DAGGERSHORT, the proof of Theorem 1 (FDR control for \DAGGERSHORT), and additional experiments.

\bibliographystyle{biometrika}

{
\bibliography{DAG}

\begin{thebibliography}{37}
\expandafter\ifx\csname natexlab\endcsname\relax\def\natexlab#1{#1}\fi

\bibitem[{Aharoni \& Rosset(2014)}]{aharoni2014generalized}
\textsc{Aharoni, E.} \& \textsc{Rosset, S.} (2014).
\newblock Generalized $\alpha$-investing: definitions, optimality results and
  application to public databases.
\newblock \textit{Journal of the Royal Statistical Society: Series B
  (Statistical Methodology)} \textbf{76}, 771--794.

\bibitem[{Ashburner et~al.(2000)Ashburner, Ball, Blake, Botstein, Butler,
  Cherry, Davis, Dolinski, Dwight, Eppig et~al.}]{ashburner2000gene}
\textsc{Ashburner, M.}, \textsc{Ball, C.~A.}, \textsc{Blake, J.~A.},
  \textsc{Botstein, D.}, \textsc{Butler, H.}, \textsc{Cherry, J.~M.},
  \textsc{Davis, A.~P.}, \textsc{Dolinski, K.}, \textsc{Dwight, S.~S.},
  \textsc{Eppig, J.~T.} et~al. (2000).
\newblock Gene ontology: tool for the unification of biology.
\newblock \textit{Nature Genetics} \textbf{25}, 25--29.

\bibitem[{Barber et~al.(2015)Barber, Cand{\`e}s et~al.}]{knockoffs}
\textsc{Barber, R.~F.}, \textsc{Cand{\`e}s, E.~J.} et~al. (2015).
\newblock Controlling the false discovery rate via knockoffs.
\newblock \textit{The Annals of Statistics} \textbf{43}, 2055--2085.

\bibitem[{Benjamini \& Hochberg(1995)}]{BH95}
\textsc{Benjamini, Y.} \& \textsc{Hochberg, Y.} (1995).
\newblock Controlling the false discovery rate: a practical and powerful
  approach to multiple testing.
\newblock \textit{Journal of the Royal Statistical Society: Series B
  (Statistical Methodology)} \textbf{57}, 289--300.

\bibitem[{Benjamini \& Yekutieli(2001)}]{BY01}
\textsc{Benjamini, Y.} \& \textsc{Yekutieli, D.} (2001).
\newblock The control of the false discovery rate in multiple testing under
  dependency.
\newblock \textit{The Annals of Statistics} \textbf{29}, 1165--1188.

\bibitem[{Blanchard \& Roquain(2008)}]{blanchard2008two}
\textsc{Blanchard, G.} \& \textsc{Roquain, E.} (2008).
\newblock Two simple sufficient conditions for {FDR} control.
\newblock \textit{Electronic Journal of Statistics} \textbf{2}, 963--992.

\bibitem[{Foster \& Stine(2008)}]{FS08}
\textsc{Foster, D.~P.} \& \textsc{Stine, R.~A.} (2008).
\newblock $\alpha$-investing: a procedure for sequential control of expected
  false discoveries.
\newblock \textit{Journal of the Royal Statistical Society: Series B
  (Statistical Methodology)} \textbf{70}, 429--444.

\bibitem[{Goeman \& Mansmann(2008)}]{goeman2008multiple}
\textsc{Goeman, J.~J.} \& \textsc{Mansmann, U.} (2008).
\newblock Multiple testing on the directed acyclic graph of gene ontology.
\newblock \textit{Bioinformatics} \textbf{24}, 537--544.

\bibitem[{Golub(2016)}]{golub2016data}
\textsc{Golub, T.} (2016).
\newblock \textit{golubEsets: exprSets for golub leukemia data}.
\newblock R package version 1.16.0.

\bibitem[{Golub et~al.(1999)Golub, Slonim, Tamayo, Huard, Gaasenbeek, Mesirov,
  Coller, Loh, Downing, Caligiuri et~al.}]{golub1999molecular}
\textsc{Golub, T.~R.}, \textsc{Slonim, D.~K.}, \textsc{Tamayo, P.},
  \textsc{Huard, C.}, \textsc{Gaasenbeek, M.}, \textsc{Mesirov, J.~P.},
  \textsc{Coller, H.}, \textsc{Loh, M.~L.}, \textsc{Downing, J.~R.},
  \textsc{Caligiuri, M.~A.} et~al. (1999).
\newblock Molecular classification of cancer: class discovery and class
  prediction by gene expression monitoring.
\newblock \textit{Science} \textbf{286}, 531--537.

\bibitem[{G'Sell et~al.(2016)G'Sell, Wager, Chouldechova \&
  Tibshirani}]{g2013false}
\textsc{G'Sell, M.~G.}, \textsc{Wager, S.}, \textsc{Chouldechova, A.} \&
  \textsc{Tibshirani, R.} (2016).
\newblock Sequential selection procedures and false discovery rate control.
\newblock \textit{Journal of the Royal Statistical Society: Series B
  (Statistical Methodology)} \textbf{78}, 423--444.

\bibitem[{Hummel et~al.(2008)Hummel, Meister \&
  Mansmann}]{doi:10.1093/bioinformatics/btm531}
\textsc{Hummel, M.}, \textsc{Meister, R.} \& \textsc{Mansmann, U.} (2008).
\newblock Globalancova: exploration and assessment of gene group effects.
\newblock \textit{Bioinformatics} \textbf{24}, 78.

\bibitem[{Hummel et~al.(2016)Hummel, Meister \& Mansmann}]{hummel2016global}
\textsc{Hummel, M.}, \textsc{Meister, R.} \& \textsc{Mansmann, U.} (2016).
\newblock Global testing of differential gene expression.
\newblock \textit{Changes} \textbf{1}, 2.

\bibitem[{Javanmard \& Montanari(2018)}]{JM16}
\textsc{Javanmard, A.} \& \textsc{Montanari, A.} (2018).
\newblock Online rules for control of false discovery rate and false discovery
  exceedance.
\newblock \textit{The Annals of Statistics} \textbf{46}, 526--554.

\bibitem[{Karlin \& Rinott(1980)}]{karlin1980classes}
\textsc{Karlin, S.} \& \textsc{Rinott, Y.} (1980).
\newblock Classes of orderings of measures and related correlation
  inequalities. {I}. multivariate totally positive distributions.
\newblock \textit{Journal of Multivariate Analysis} \textbf{10}, 467--498.

\bibitem[{Katsevich \& Ramdas(2018)}]{katsevich2018towards}
\textsc{Katsevich, E.} \& \textsc{Ramdas, A.} (2018).
\newblock Towards" simultaneous selective inference": post-hoc bounds on the
  false discovery proportion.
\newblock \textit{arXiv preprint arXiv:1803.06790} .

\bibitem[{Lehmann(1966)}]{lehmann1966some}
\textsc{Lehmann, E.~L.} (1966).
\newblock Some concepts of dependence.
\newblock \textit{The Annals of Mathematical Statistics} \textbf{37},
  1137--1153.

\bibitem[{Lei et~al.(2017)Lei, Ramdas \& Fithian}]{lei17star}
\textsc{Lei, L.}, \textsc{Ramdas, A.} \& \textsc{Fithian, W.} (2017).
\newblock {STAR}: a general interactive framework for {FDR} control under
  structural constraints.
\newblock \textit{arXiv preprint arXiv:1710.02776} .

\bibitem[{Li \& Barber(2017)}]{li2017accumulation}
\textsc{Li, A.} \& \textsc{Barber, R.~F.} (2017).
\newblock Accumulation tests for {FDR} control in ordered hypothesis testing.
\newblock \textit{Journal of the American Statistical Association}
  \textbf{112}, 837--849.

\bibitem[{Lynch(2014)}]{lynch2014control}
\textsc{Lynch, G.} (2014).
\newblock \textit{The Control of the False Discovery Rate Under Structured
  Hypotheses}.
\newblock Ph.D. thesis, New Jersey Institute of Technology, Department of
  Mathematical Sciences.

\bibitem[{Lynch \& Guo(2016)}]{lynch2016procedures}
\textsc{Lynch, G.} \& \textsc{Guo, W.} (2016).
\newblock On procedures controlling the {FDR} for testing hierarchically
  ordered hypotheses.
\newblock \textit{arXiv preprint arXiv:1612.04467} .

\bibitem[{Lynch et~al.(2016)Lynch, Guo, Sarkar \& Finner}]{lynch2016control}
\textsc{Lynch, G.}, \textsc{Guo, W.}, \textsc{Sarkar, S.~K.} \& \textsc{Finner,
  H.} (2016).
\newblock The control of the false discovery rate in fixed sequence multiple
  testing.
\newblock \textit{arXiv preprint arXiv:1611.03146} .

\bibitem[{Meijer \& Goeman(2015{\natexlab{a}})}]{meijer2015multiple1}
\textsc{Meijer, R.~J.} \& \textsc{Goeman, J.~J.} (2015{\natexlab{a}}).
\newblock A multiple testing method for hypotheses structured in a directed
  acyclic graph.
\newblock \textit{Biometrical Journal} \textbf{57}, 123--143.

\bibitem[{Meijer \& Goeman(2015{\natexlab{b}})}]{meijer2015multiple2}
\textsc{Meijer, R.~J.} \& \textsc{Goeman, J.~J.} (2015{\natexlab{b}}).
\newblock Multiple testing of gene sets from gene ontology: possibilities and
  pitfalls.
\newblock \textit{Briefings in bioinformatics} \textbf{17}, 808--818.

\bibitem[{Meinshausen(2008)}]{meinshausen2008hierarchical}
\textsc{Meinshausen, N.} (2008).
\newblock Hierarchical testing of variable importance.
\newblock \textit{Biometrika} \textbf{95}, 265--278.

\bibitem[{Ramdas et~al.(2018{\natexlab{a}})Ramdas, Barber, Wainwright \&
  Jordan}]{ramdas2017unified}
\textsc{Ramdas, A.}, \textsc{Barber, R.~F.}, \textsc{Wainwright, M.~J.} \&
  \textsc{Jordan, M.~I.} (2018{\natexlab{a}}).
\newblock A unified treatment of multiple testing with prior knowledge using
  the p-filter.
\newblock \textit{The Annals of Statistics (accepted)} .

\bibitem[{Ramdas et~al.(2017)Ramdas, Yang, Wainwright \& Jordan}]{RYWJ17}
\textsc{Ramdas, A.}, \textsc{Yang, F.}, \textsc{Wainwright, M.~J.} \&
  \textsc{Jordan, M.~I.} (2017).
\newblock Online control of the false discovery rate with decaying memory.
\newblock In \textit{Advances In Neural Information Processing Systems}.

\bibitem[{Ramdas et~al.(2018{\natexlab{b}})Ramdas, Zrnic, Wainwright \&
  Jordan}]{SAFFRON}
\textsc{Ramdas, A.}, \textsc{Zrnic, T.}, \textsc{Wainwright, M.~J.} \&
  \textsc{Jordan, M.~I.} (2018{\natexlab{b}}).
\newblock {SAFFRON}: an adaptive algorithm for online control of the false
  discovery rate.
\newblock In \textit{35th International Conference on Machine Learning}.

\bibitem[{Rosenbaum(2008)}]{rosenbaum2008testing}
\textsc{Rosenbaum, P.~R.} (2008).
\newblock Testing hypotheses in order.
\newblock \textit{Biometrika} \textbf{95}, 248--252.

\bibitem[{R{\"u}ger(1978)}]{ruger1978maximale}
\textsc{R{\"u}ger, B.} (1978).
\newblock Das maximale signifikanzniveau des tests: ``{L}ehne ${H}_0$ ab, wenn
  $k$ unter $n$ gegebenen tests zur ablehnung f{\"u}hren''.
\newblock \textit{Metrika} \textbf{25}, 171--178.

\bibitem[{R{\"u}schendorf(1982)}]{ruschendorf1982random}
\textsc{R{\"u}schendorf, L.} (1982).
\newblock Random variables with maximum sums.
\newblock \textit{Advances in Applied Probability} , 623--632.

\bibitem[{Saunders et~al.(2014)Saunders, Stevens \&
  Isom}]{saunders2014shortcut}
\textsc{Saunders, G.}, \textsc{Stevens, J.~R.} \& \textsc{Isom, S.~C.} (2014).
\newblock A shortcut for multiple testing on the directed acyclic graph of gene
  ontology.
\newblock \textit{BMC Bioinformatics} \textbf{15}, 349.

\bibitem[{Simes(1986)}]{Simes1986improved}
\textsc{Simes, J.} (1986).
\newblock An improved {B}onferroni procedure for multiple tests of
  significance.
\newblock \textit{Biometrika} \textbf{73}, 751--754.

\bibitem[{Storey et~al.(2004)Storey, Taylor \& Siegmund}]{Storey04}
\textsc{Storey, J.}, \textsc{Taylor, J.} \& \textsc{Siegmund, D.} (2004).
\newblock Strong control, conservative point estimation and simultaneous
  conservative consistency of false discovery rates: a unified approach.
\newblock \textit{Journal of the Royal Statistical Society: Series B
  (Statistical Methodology)} \textbf{66}, 187--205.

\bibitem[{Stouffer et~al.(1949)Stouffer, Suchman, DeVinney, Star \&
  Williams~Jr}]{stouffer1949american}
\textsc{Stouffer, S.~A.}, \textsc{Suchman, E.~A.}, \textsc{DeVinney, L.~C.},
  \textsc{Star, S.~A.} \& \textsc{Williams~Jr, R.~M.} (1949).
\newblock The {A}merican soldier: {A}djustment during army life, vol. 1 .

\bibitem[{Vovk(2012)}]{vovk2012combining}
\textsc{Vovk, V.} (2012).
\newblock Combining p-values via averaging.
\newblock \textit{arXiv preprint arXiv:1212.4966} .

\bibitem[{Yekutieli(2008)}]{yekutieli2008hierarchical}
\textsc{Yekutieli, D.} (2008).
\newblock Hierarchical false discovery rate--controlling methodology.
\newblock \textit{Journal of the American Statistical Association}
  \textbf{103}, 309--316.

\end{thebibliography}
}

\newpage
\appendix

\section{An illustrative example of \DAGGER}

\label{sec:example}
\begin{table}[!ht] 
\centering
\resizebox{0.80\linewidth}{!}{
\begin{tabular}{||c|c|c||} 
 \hline
Graph & $m$ and $l$ & Steps of \DAGGER \\ [0.5ex] 
\hline
\image{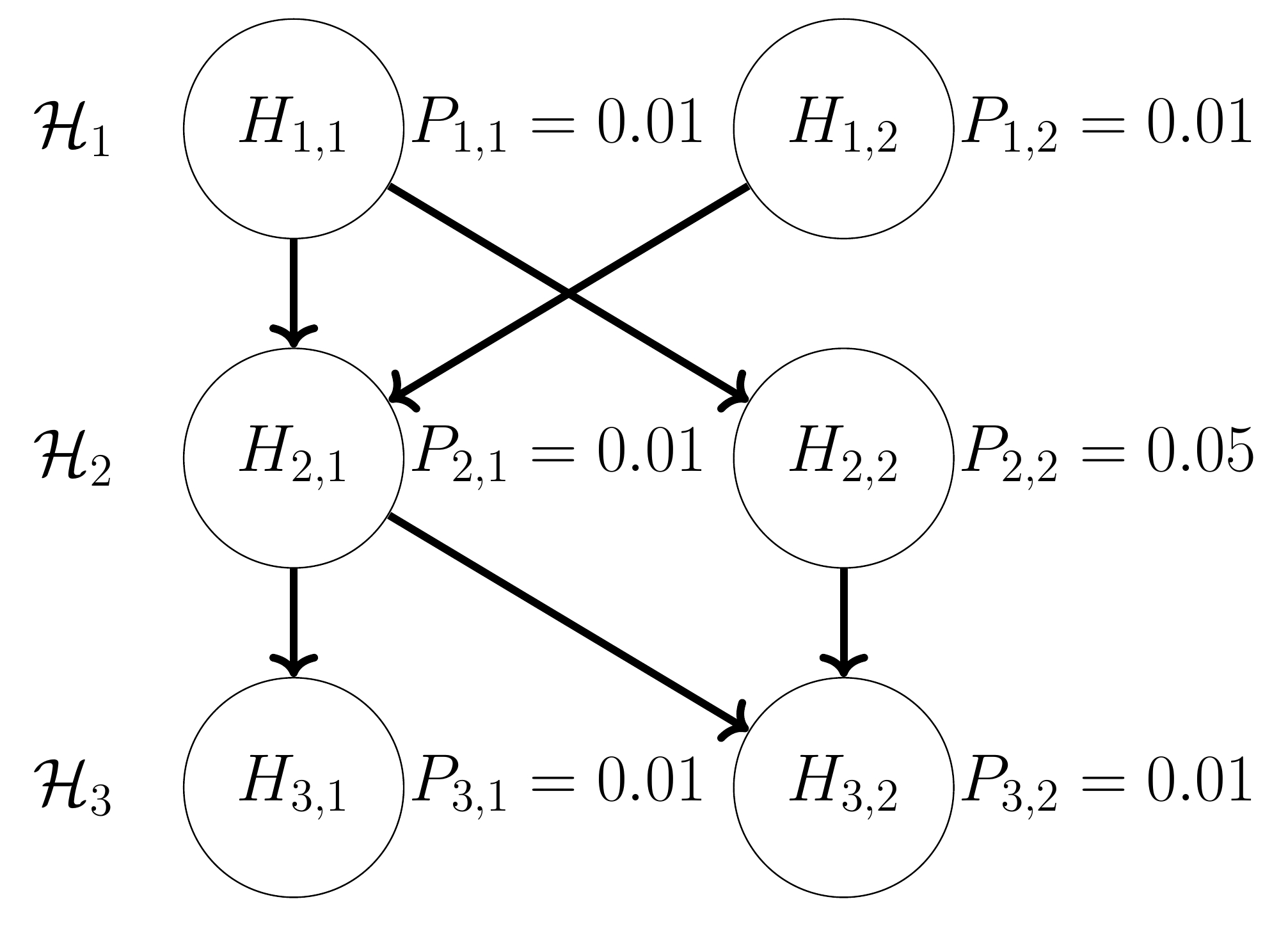} & 
& \\
 \hline
 \image{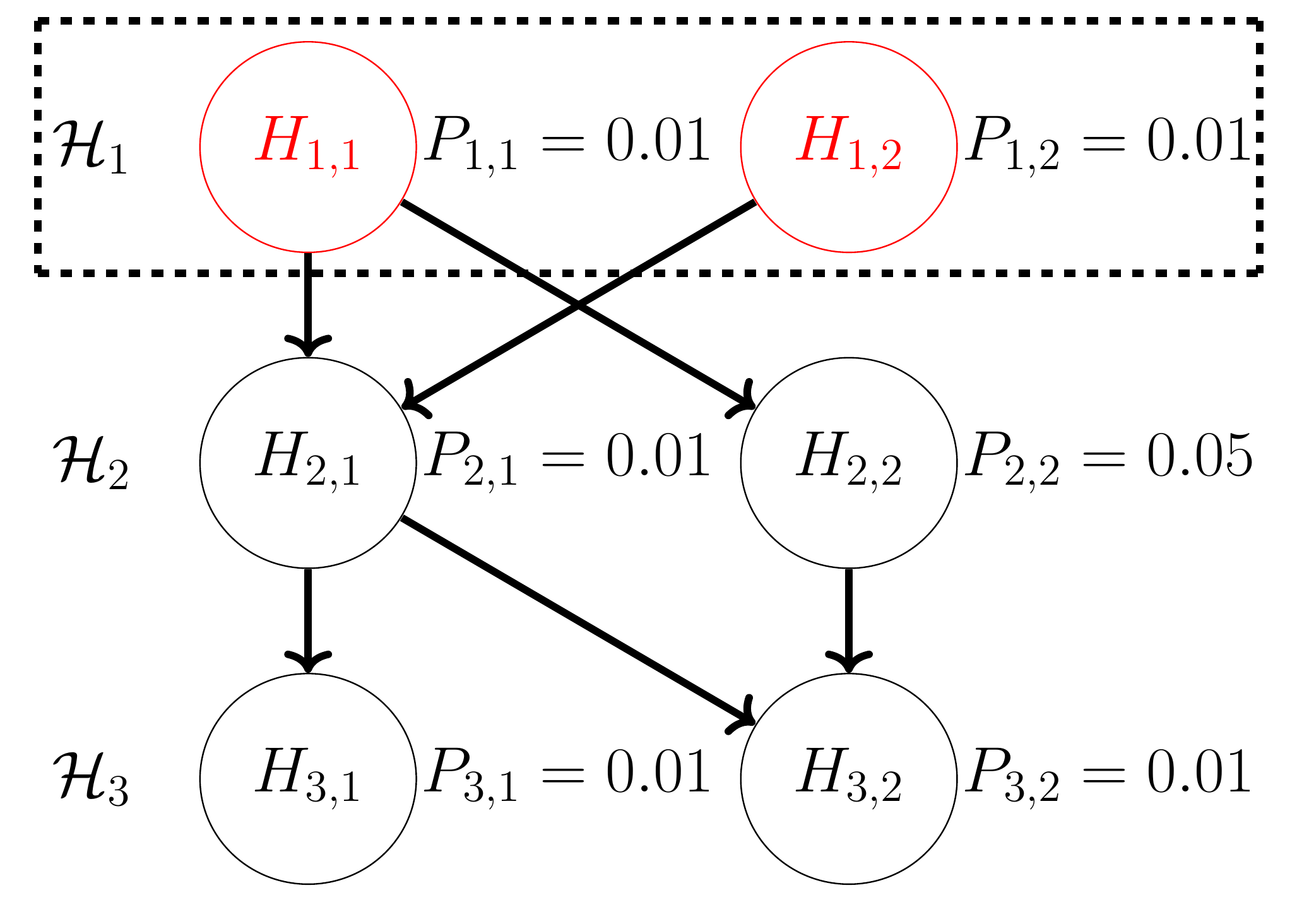} & 
\makecell{$m_{1,1} = 1+0.5m_{2,1}+m_{2,2}=3.75$,\\
$m_{1,2}=1+0.5m_{2,1}=2.25$,
\vspace{2cm}\\
$l_{1,1}=0.5l_{2,1}+l_{2,2}=1.25$,\\
$l_{1,2}=0.5l_{2,1}=0.75$.} & 
\makecell{$\alpha_{1,1}(r)=\frac{1.25}{2}\frac{3.75+r-1}{3.75}\alpha,$ 
\vspace{0.2cm}\\
$\alpha_{1,2}(r)=\frac{0.75}{2}\frac{2.25+r-1}{2.25}\alpha,$ 
\vspace{0.2cm}\\ $P_{1,1}\leq \alpha_{1,1}(2)$ and $P_{1,2}\leq
\alpha_{1,2}(2).$\\ Plugging into equation (7), \\ we find that both
nodes are rejected.}\\ \hline \image{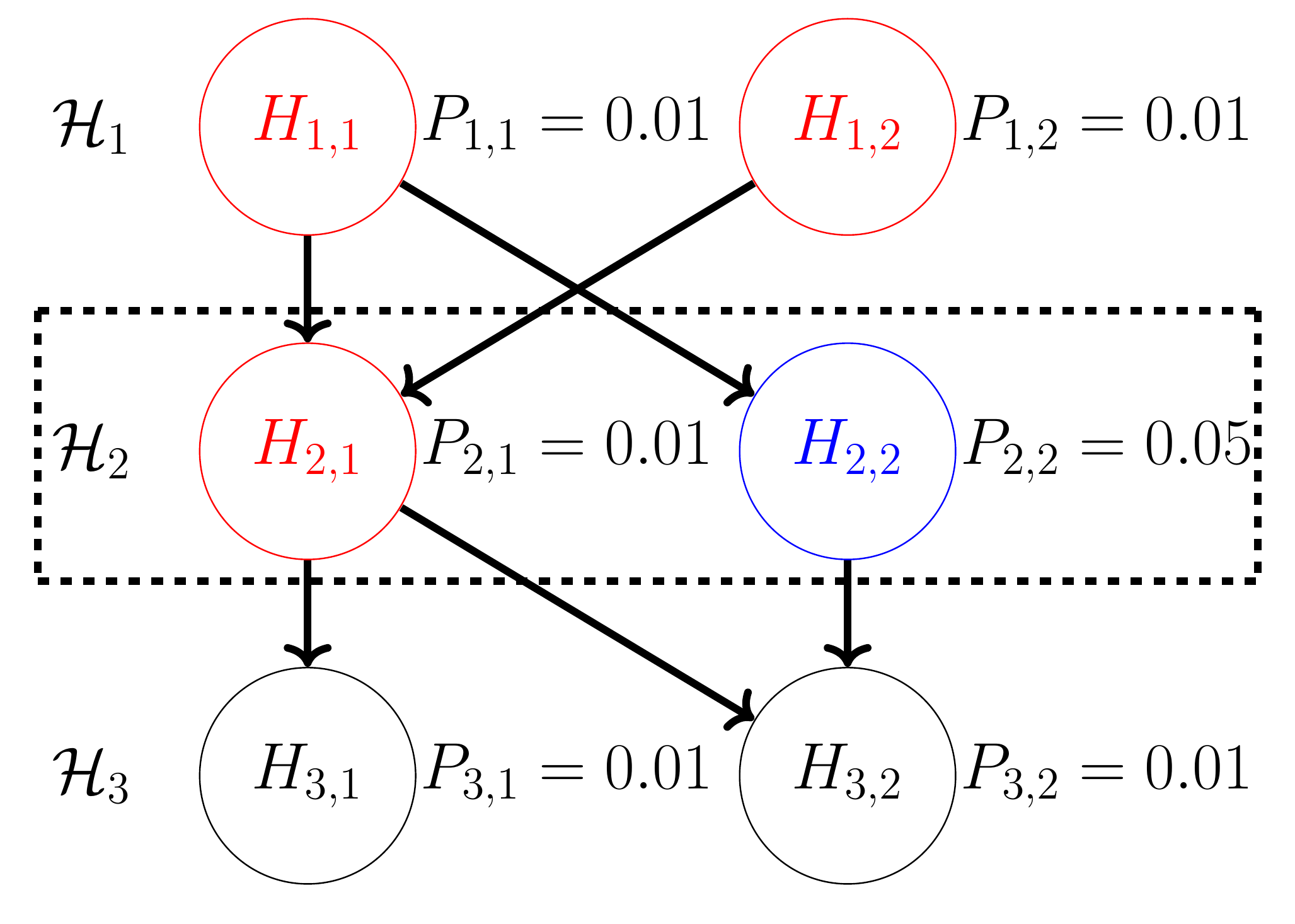} &
\makecell{$m_{2,1}=1+m_{3,1}+0.5m_{3,2}=2.5$,\\ $m_{2,2}=1+0.5m_{3,2}=1.5$,
\vspace{2cm}\\
$l_{2,1}=l_{3,1}+0.5l_{3,2}=1.5$,\\
$l_{2,2}=0.5l_{3,2}=0.5$.} & 
\makecell{$\alpha_{2,1}(r)=\frac{1.5}{2}\frac{2.5+r+2-1}{2.5}\alpha,$ 
\vspace{0.2cm}\\
$\alpha_{2,2}(r)=\frac{0.5}{2}\frac{1.5+r+2-1}{1.5}\alpha,$ 
\vspace{0.2cm}\\ $P_{2,1} \leq \alpha_{2,1}(2)$ but $P_{2,2} >
\alpha_{2,2}(2),$\\ $P_{2,1} \leq \alpha_{2,1}(1)$ and $P_{2,2} >
\alpha_{2,2}(1).$\\ Plugging into equation (7), \\ we find that
$R_2=1$,\\ and only $H_{2,1}$ is rejected.}\\ \hline
\image{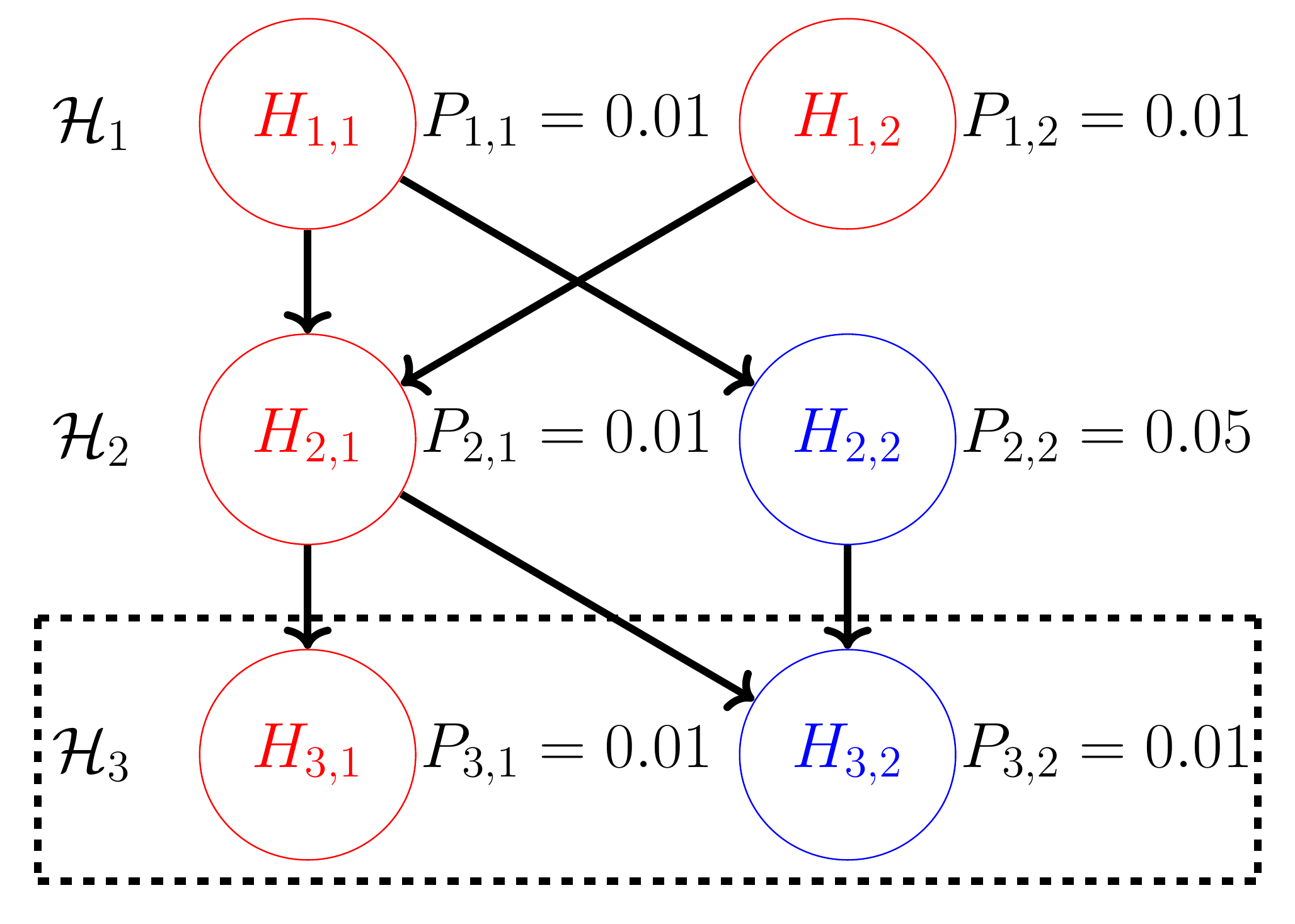} & \makecell{$m_{3,1}=1$,\\$m_{3,2}=1$,
\vspace{2cm}\\
$l_{3,1}=1$,\\
$l_{3,2}=1$,\\
$L=2$.} & 
\makecell{$H_{3,2}$ is not tested,\\
as one of its parents is not rejected.\\
$\alpha_{3,1}(r)=\frac 1 2 (1+r+3-1)\alpha,$\\
$P_{3,1}\leq \alpha_{3,1}(1),$\\
so only $H_{3,1}$ is rejected.
}\\
 \hline   
\end{tabular}
}
  \caption{
An illustrative example of how \DAGGER works on a toy DAG with independent $p$-values. The target FDR level is $\alpha=0.05$. The first row is the DAG with $p$-values to be considered. The following three rows show progressive steps of \DAGGER layer by layer, with the focused level in a dashed box. The nodes rejected, not rejected, and not yet tested are in red, blue, and black respectively. The second column shows the calculation of effective numbers of nodes and leaves in a bottom-up fashion, and the third column explicitly spells out calculations done by \DAGGER.}
\label{table:supp1}
\end{table}



\newpage
\section{The proof of Theorem 1 (FDR control for \DAGGER)}\label{sec:proofs}

Below, we will prove that $\fdr(\H_{1:D}) = \fdr \leq \alpha$, from which it must necessarily follow that $\fdr(\H_{1:d}) \leq \alpha$ for all $d$. The latter implication is true because \DAGGER must control FDR regardless of null $p$-values, and when it has completed $d$ rounds (for any $d$), it must automatically guard against the hypothetical possibility that all remaining hypotheses are nulls whose $p$-values equal 1, in which case \DAGGER will make no more rejections, resulting in $\fdr(\H_{1:d})$ equaling $\fdr(\H_{1:D})$ and hence being at most $\alpha$. 

To proceed, we first define random variables that we need in the proofs, namely the effective number of discoveries and false discoveries in the subgraph below a node. Recalling that $\Rejections$ and $\False$  are the set of discoveries and false discoveries, we define $R(\Subgraph(a))$ and $V(\Subgraph(a))$ in a bottom-up fashion. They are set for leaves $a \in \Leaves$ as
\begin{align*}
V(\Subgraph(a)) &:= \One{a \in \False},\\
R(\Subgraph(a)) &:= \One{a \in \Rejections}.
\end{align*}
We then calculate their effective values for non-leaf nodes $a$ recursively up the DAG as
\begin{align}
V(\Subgraph(a)) &~:=~ \One{a \in \False} + \sum_{b \in \Children(a)} \frac{V(\Subgraph(b))}{|\Parents(b)|},\label{eqn:recursefalse}\\
R(\Subgraph(a)) &~:=~ \One{a \in \Rejections} + \sum_{b \in \Children(a)} \frac{R(\Subgraph(b))}{|\Parents(b)|}. \nonumber
\end{align}
They satisfy the following identities at the roots:
\begin{align}
V &= \sum_{a \in \H_1} V(\Subgraph(a)) \text{~ and ~} R ~= \sum_{a \in \H_1} R(\Subgraph(a)). \label{eqn:identityfalse}
\end{align}

\noindent Before proceeding, we summarize the notation used in the proof.
 
 \vspace{0.1in}
\resizebox{0.95\linewidth}{!}{
\begin{tabular}{ | l l | }
\hline 
Notation & Meaning \\ 
\hline 
$N$ & the number of nodes in the DAG, each one representing a null\\
$D$ & the maximum depth of any node in the DAG\\
$\H_d, d \in \{1,\dots,D\}$ & the set of hypotheses at depth $d$ \\
$\H_{1:d}$ & the set of hypotheses with depth less than or equal to  $d$\\
$H_{d,i}, d \in \{1,\dots,D\}, i\in \{1,\dots,|\H_d|\}$ & the $N$ null hypotheses, one at each node\\
$\nulls$ & the set of all null hypotheses \\
$P_{d,i}$ & the $N$ $p$-values, one at each node\\
$\Leaves$ & the set of all leaves in the DAG\\
$L = |\Leaves|$ & the total number of leaves in the DAG\\
\hline
$\alpha_{d,i}$ & the level at which $H_{d,i}$ is tested\\
$R_{d,i}$ & the indicator of whether the $i$-th hypothesis was rejected\\
$\Rejections$ & the set of all rejected hypotheses \\
$R = \sum_{d,i} R_{d,i}$ & the total number of rejections\\
$\Rejections(\H_{1:d})$ & the set of all rejections up until, and including depth $d$\\
$R(\H_{1:d})$ & the total number of rejections up until, and including depth $d$\\
$\False = \nulls \cap \Rejections$ & the set of all false discoveries \\
$V$ & the total number of false discoveries\\
\hline
$\Depth(a)$ & the depth of node $a$ \\
$\Parents(a)$ & the set of all parents of node $a$\\
$\Children(a)$ & the set of all children of node $a$\\
$\Subgraph(a)$ & the set of all descendents of node $a$, including node $a$\\ 
\hline
$m_a \leq |\Subgraph(a)|$ & the \textit{effective} number of nodes in $\Subgraph(a)$\\
$\ell_a $ & the \textit{effective} number of leaves in $\Subgraph(a)$\\
$R(\Subgraph(a)) \leq m_a$ & the \textit{effective} number of discoveries made in $\Subgraph(a)$ \\
$V(\Subgraph(a))$ & the \textit{effective} number of false discoveries made in $\Subgraph(a)$\\
\hline
\end{tabular}
}


\noindent The key enabler of the proof of Theorem~1 is the following lemma. 

\begin{lemma}\label{lem:subFDR}
Under the assumptions of Theorem~1, \DAGGER guarantees that
\begin{align}
  \EE{\frac{V(\Subgraph(a))}{R}} \leq \alpha \frac{\ell_a}{L} \qquad
  \mbox{for all  nodes $a$ in the DAG.}
\end{align}
\end{lemma}
Note that this lemma implies the statement of the theorem. 
Indeed, we have
\begin{align*}
\fdr = \EE{\frac{V}{R}} &= \sum_{a \in \H_1}
\EE{\frac{V(\Subgraph(a))}{R}} \\ &\leq \sum_{a \in \H_1} \alpha
\frac{\ell_a}{L}\\ &= \alpha,
\end{align*}
where the first equality follows by identity~\eqref{eqn:identityfalse}, the inequality by \lemref{subFDR} and the last equality follows from the definition of effective number of leaves.

Hence, to conclude the proof of Theorem~1, we only need to prove \lemref{subFDR} under the different dependence assumptions. With this aim in mind, we utilize variants of the super-uniformity lemmas proved by \citet{ramdas2017unified}.  In particular, first note that our super-uniformity assumption on null $p$-values can be reformulated as follows:
\begin{align}
\label{eqn:PRDS-fixed}
\text{For any $i \in \nulls$, ~}~ \EE{\frac{{\One{P_i\leq t}}}{t}} & \leq 1  \text{ for any
   non-random $t \in [0,1]$.}
\end{align}
Of course, if $P_i$ is uniform then the above inequality holds with
equality.  The following lemma, which combines parts of Lemmas 1 and 3
from \citet{ramdas2017unified}, guarantees that
property~\eqnref{PRDS-fixed} continues to hold for certain random
thresholds $f(P)$.  Recall that the term ``nonincreasing'' is
interpreted coordinatewise, with respect to the orthant ordering.
\begin{lemma}[Super-uniformity lemmas \citep{ramdas2017unified}]
\label{lem:power}
Let $f :[0,1]^n\mapsto [0,\infty)$ be an arbitrary function, let index $i \in
  \nulls$ refer to some null hypothesis, and let  $g \subseteq \nulls$ refer to some null group with Simes' $p$-value $P_g$.
  \begin{enumerate}
    \addtolength{\itemindent}{0.6cm}%
    \item[(a)] If  $f$ is nonincreasing, then under independence or positive dependence, we have
\[
\EE{\frac{\One{P_i\leq f(P)}}{f(P)} } \leq 1.
\]
\item[(b)] For any constant $c>0$ and $\beta \in \beta(\cT)$, under arbitrary dependence we have
\[
\EE{\frac{\One{P_i\leq c \beta(f(P)) )}}{cf(P)} } \leq 1 .
\]
 \item[(c)] If $f$ is nonincreasing, and the base $p$-values are positively dependent, we have
\[
\EE{\frac{\One{P_g \leq f(P)}}{f(P)} } \leq 1.
\]
\end{enumerate}
\end{lemma} 

We will apply the above super-uniformity lemma to prove \lemref{subFDR}, where cases (a,b,c) above will correspond exactly to cases (a,b,c) in the statement of Theorem~1. Theorem statement (d) actually follows directly from theorem statement (b)---independence at leaf $p$-values is only required so that $p$-value combination methods like Fisher's and Rosenthal's can produce valid $p$-values; however, since different Fisher combinations can be arbitrarily dependent, we may utilize reshaping to protect against this scenario. 

We now have the tools in place to prove \lemref{subFDR}, concluding
the proof of Theorem~1. At a high level, \lemref{subFDR} is
proved using a bottom-up induction.

\begin{proof}

We divide the proof into two cases, depending on whether or not $H_a
\equiv H_{d,i}$.


Case 1: Suppose $H_{d,i}$ is true. Note that
$V(\Subgraph(a))$ is nonzero only if $H_{d,i}$ was rejected. Since
$V(\Subgraph(a)) \leq R(\Subgraph(a)) \leq m_a$, we have
\begin{align*}
\EE{\frac{V(\Subgraph(a))}{R}} &\leq
\EE{\frac{V(\Subgraph(a))}{V(\Subgraph(a)) + R - R(\Subgraph(a))}}
\\ &\leq \EE{\frac{m_a }{m_a + R - R(\Subgraph(a))} \One{H_{d,i}
    \text{ is rejected }} } \\ &\leq \EE{\frac{m_a }{m_a + R(\H_{1:d})
    - 1} \One{H_{d,i} \text{ is rejected }} },
\end{align*}
where the last inequality follows because $R = R(\H_{1:D}) \geq
R(\H_{1:d}) + R(\Subgraph(a)) - 1$. By definition of our procedure, we
may then infer that
\begin{align*}
\EE{\frac{V(\Subgraph(a))}{R}} &\leq \alpha \frac{\ell_a}{L}
\EE{\frac1{\alpha}\frac{L}{\ell_a}\frac{m_a }{m_a + R(\H_{1:d}) - 1}
  \One{P_{d,i} \leq \alpha \frac{\ell_a}{L} \frac{\beta_d(m_a +
      R(\H_{1:d}) - 1)}{m_a}} }\\ &\leq \alpha \frac{\ell_a}{L},
\end{align*}
where the last inequality follows by applying \lemref{power} (a, b, c)
to the function $f: P \mapsto \alpha \frac{\ell_a}{L} \frac{m_a +
  R(\H_{1:d}) - 1}{m_a} $, which can be seen to be nonincreasing in
$P$; indeed, decreasing any of the $p$-values can only possibly
increase the total number of rejections.

Case 2: We handle the case when $H_a$ is false by induction,
leaves upwards. If $H_a$ is a leaf, the lemma is trivially true. Now
assume for the purpose of induction that the lemma is true for every
false child of $H_a$, which along with case 1 means that the lemma
would then be true for every child of $H_a$. Hence,
\begin{align*}
\EE{\frac{V(\Subgraph(a))}{R}} &= \sum_{b \in \Children(a)}
\frac1{|\Parents(b)|}\EE{\frac{V(\Subgraph(b))}{R}}\\ &\leq \sum_{b
  \in \Children(a)} \frac1{|\Parents(b)|} \alpha \frac{\ell_b}{L}\\ &
= \alpha \frac{\ell_a}{L},
\end{align*}
where the first equality follows by the
identity~\eqnref{recursefalse}, the inequality follows by the
induction hypothesis, and the last equality follows by the definition of the effective number of leaves.

\end{proof}

\noindent This concludes the proof of Theorem~1.



\section{Additional experiments to study the effects of graph structure}
\label{sec:structure}

In this appendix, we report the results of studies on how the
structure of DAGs affect the achieved FDR and power of
\DAGGERSHORT. We focus on three scenarios: In the first scenario, we
make a comparison between shallow and deep networks. In the second
scenario, we compare DAGs of two shapes: the diamond and the hourglass
DAGs. In the third scenario, we compare DAGs in the shape of a
mountain with DAGs in the shape of a valley.  In all cases, we
generate random DAGs by first determining the number of nodes at each
layer, then starting from the bottom layer, we randomly assign $k$
distinct parents at depth $d-1$ for each node at depth $d$. Given a
DAG, the leaves are randomly chosen with probability $\pi_0$ to be
true nulls. The rest of the hypotheses are assigned true if and only
if all of its children are true.


\subsection{Shallow versus deep.}
We define shallow DAGs to be DAGS composed of two layers with the same
number of nodes in each layer. Deep DAGs are composed of four layers
with the same number of nodes on each layer. For both cases, each node
in a lower layer is randomly assigned to two parents in its upper
layer. The target FDR is fixed to be $\alpha=0.2$, the number of nodes
are fixed to be $500$ and the signal strength of non-nulls is $\mu=2$.

\begin{figure}[h!]
\centering
    \includegraphics[width=.48\linewidth]{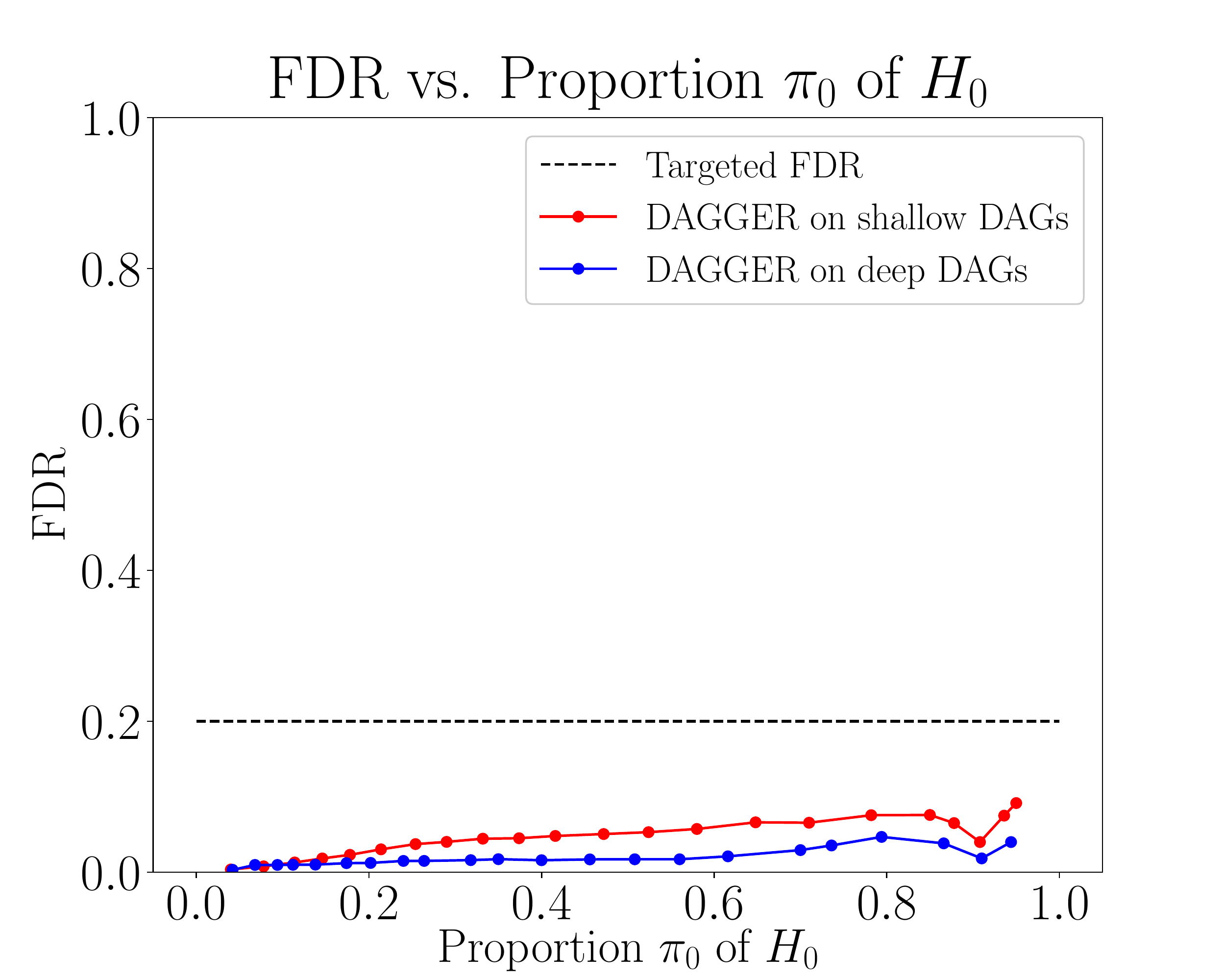}\quad 
    \includegraphics[width=.48\linewidth]{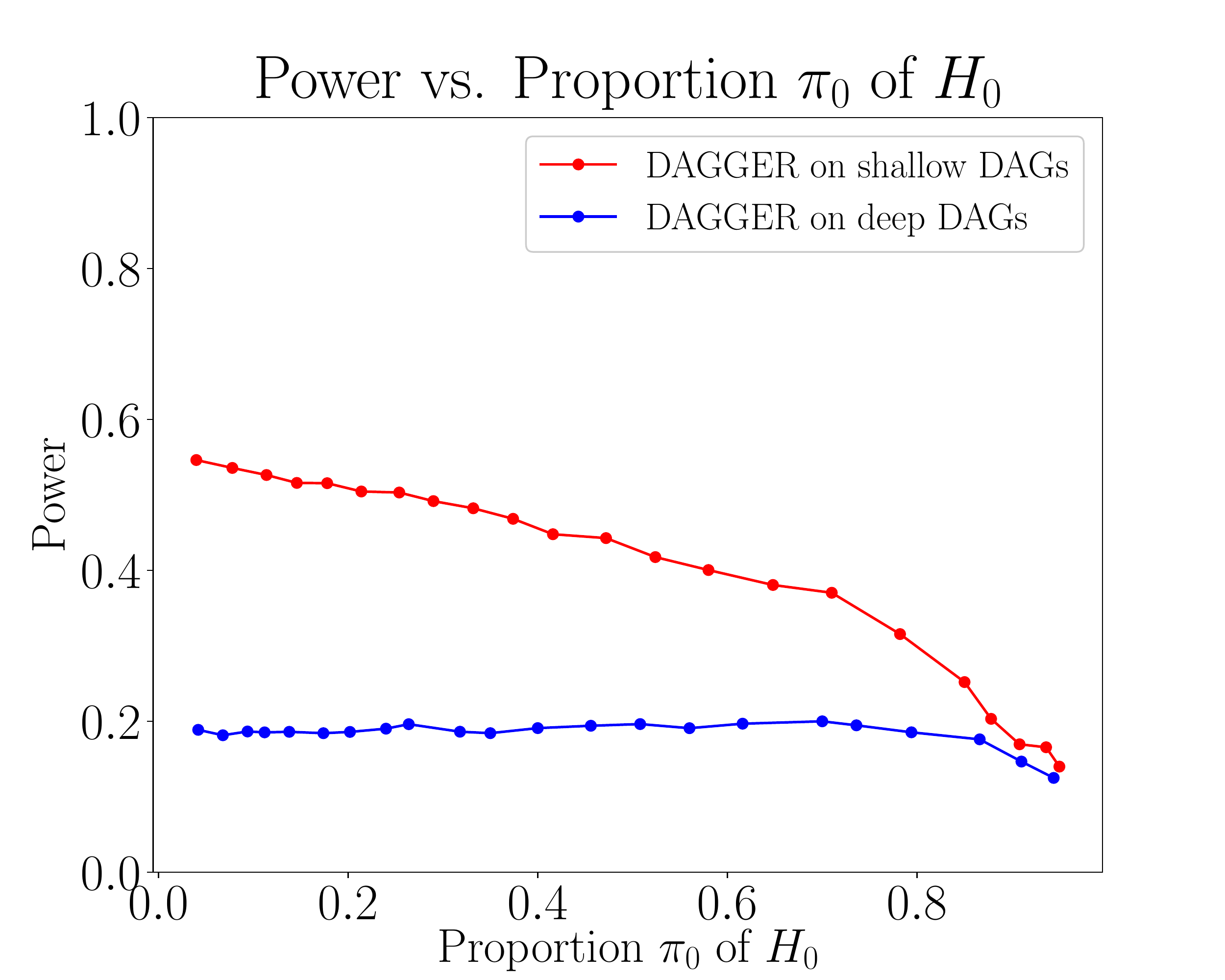}\quad
     \includegraphics[width=.48\linewidth]{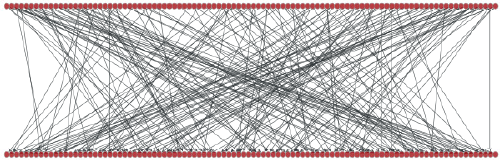}\quad 
    \includegraphics[width=.48\linewidth]{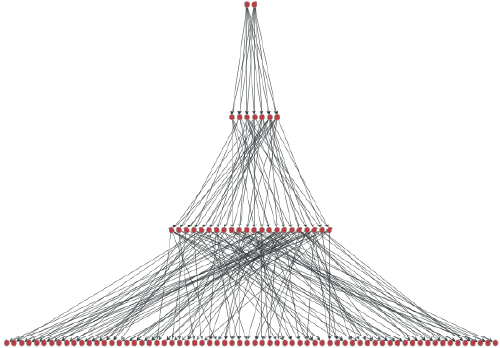}   
  \caption{Plots of the FDR and power of \DAGGER versus the proportion of nulls on shallow and deep graphs, controlling the target FDR $\alpha=0.2$, number of nodes $n=500$, signal strength $\mu=2$ and with $100$ repetitions. 
  } 
\label{fig:shallow-deep}
\end{figure}

Figure \ref{fig:shallow-deep} shows the plots of the achieved FDR and power of \DAGGER versus the proportion of nulls on shallow and deep graphs. \DAGGER achieves larger power with a higher false discovery proportion on shallow DAGs. In fact, an incorrect acceptance of a non-null in the first layer has a worse effect when the DAG is deeper, because non-nulls in its descendants have no chance to be discovered at all.

\subsection{Diamond versus hourglass.}
We define diamond  DAGs to have the largest number of nodes in middle layers and fewer nodes at the top and the bottom layers, while hourglass  DAGs have the smallest number of nodes in middle layers with more nodes at the top and bottom. 
In this experiment, we consider diamond and hourglass  DAGs with three layers and $500$ nodes in total. For diamond DAGs, there are $125, 250, 125$ nodes on the three subsequent layers. Each node in the last layer is randomly assigned to two parents in its middle layer. Each node in the middle layer is randomly assigned to one parent in the top layer. For hourglass DAGs, there are $200, 100$ and $200$ nodes on the three subsequent layers. Each node in the last layer is randomly assigned to one  parent in its middle layer. Each node in the middle layer is randomly assigned to two parents in the top layer. 

\begin{figure}[h!]
\centering
    \includegraphics[width=.45\linewidth]{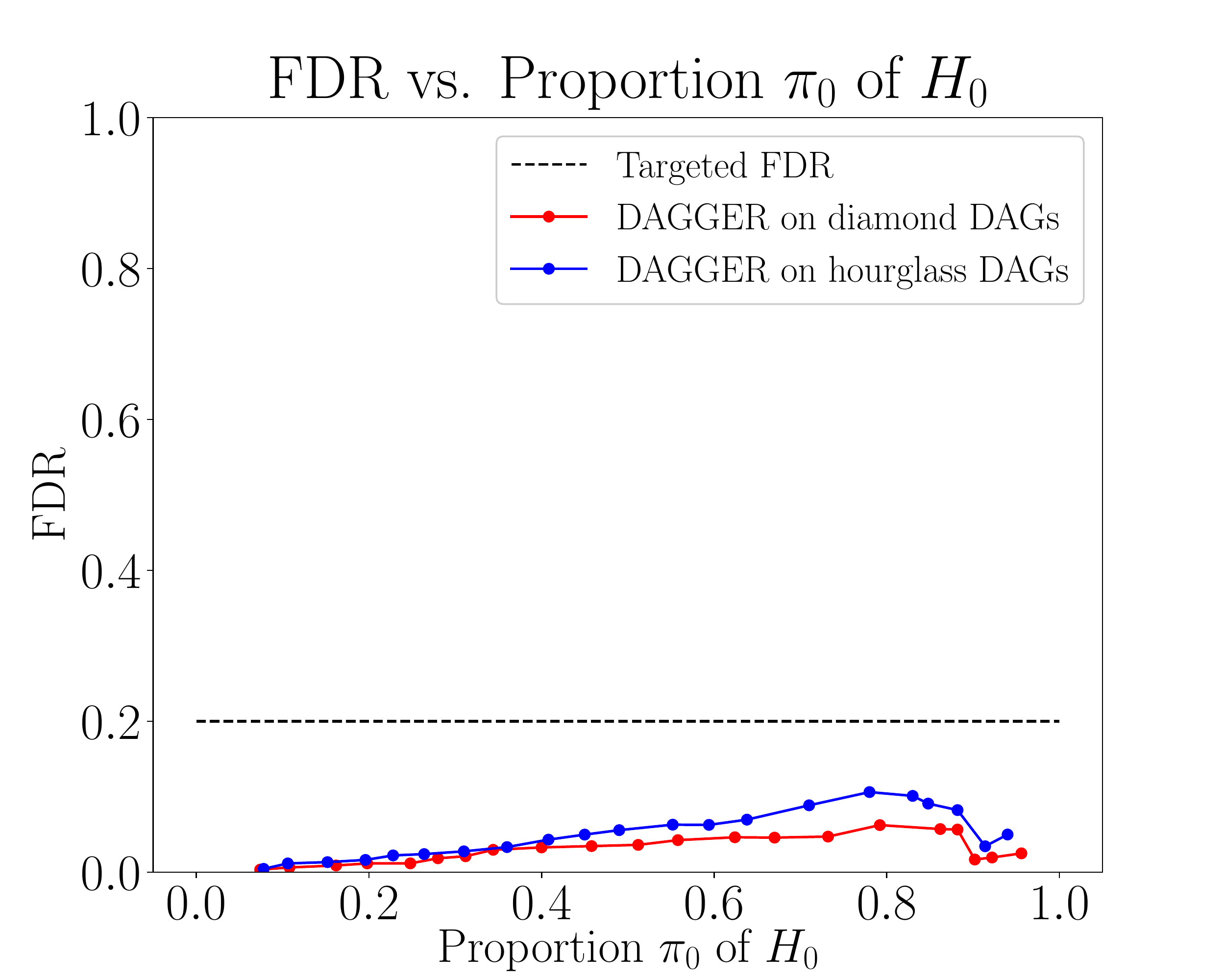}\quad 
    \includegraphics[width=.45\linewidth]{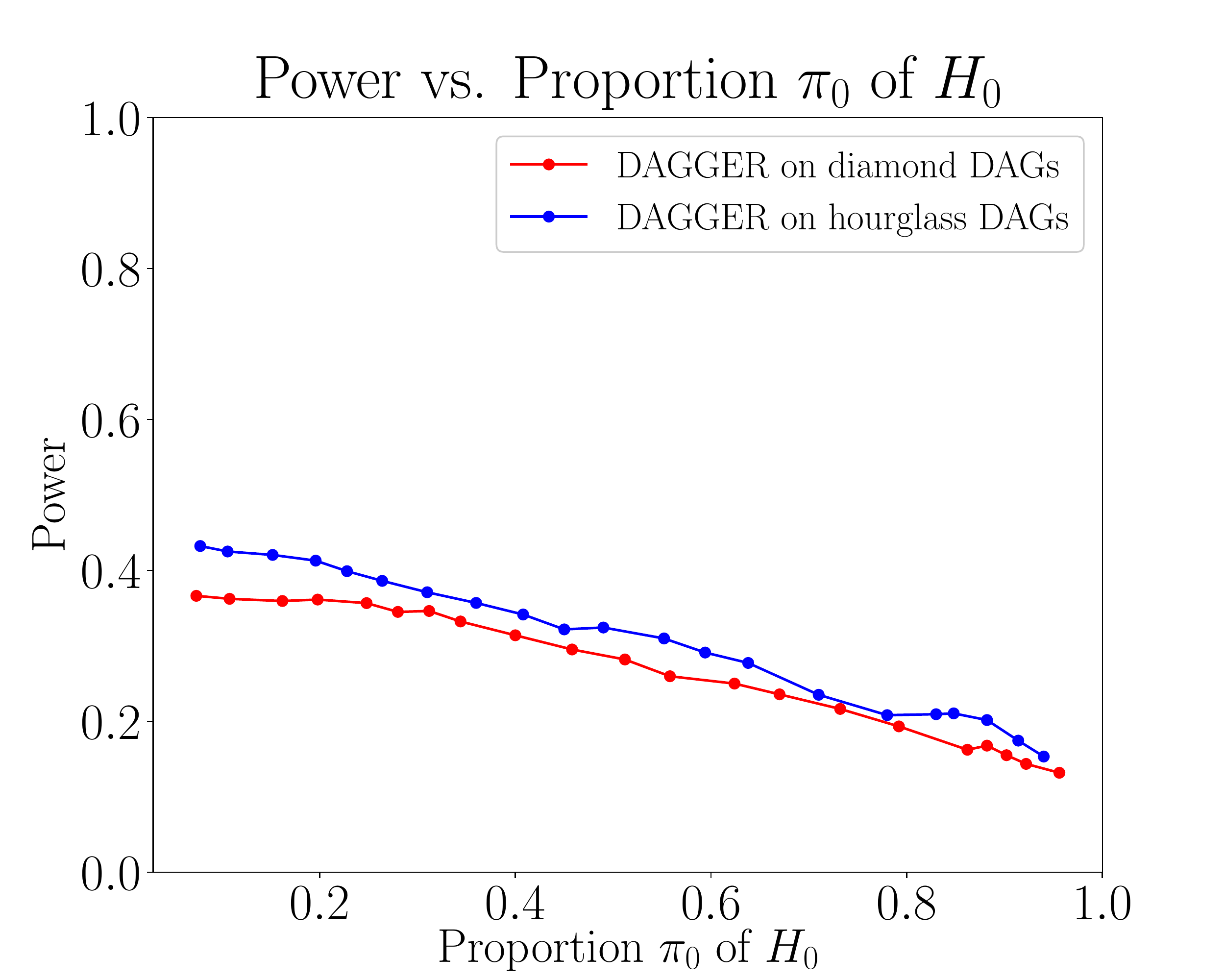} \quad
    \includegraphics[width=.45\linewidth]{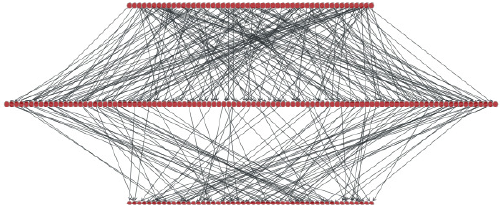}\quad 
    \includegraphics[width=.45\linewidth]{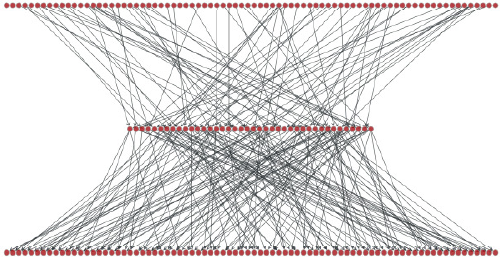}  
  \caption{Plots of the achieved FDR and power of \DAGGER versus the proportion of nulls on diamond and hourglass graphs, controlling the target FDR $\alpha=0.2$, number of nodes $n=500$, signal strength $\mu=2$ and with $100$ repetitions. 
  } 
\label{fig:diamond-hourglass}
\end{figure}

Figure \ref{fig:diamond-hourglass} shows plots of the achieved FDR and power of \DAGGER versus the proportion of nulls on diamond and hourglass graphs. We set the target FDR $\alpha=0.2$ and the signal strength of non-nulls to be $\mu=2$. \DAGGER achieves a higher power on hourglass DAGs. Intuitively, for each non-null in the top layer incorrectly accepted, there are a larger number of non-nulls ignored by \DAGGER among its descendants when the DAG is of diamond shape. For each non-null in the second layer incorrectly accepted, there are a larger number of non-nulls ignored by \DAGGER among its descendants when the DAG is of hourglass shape. Due to the structure of DAGs, the first case has a stronger negative effect on the power of DAGs, resulting in less power in diamond DAGs.

\subsection{Mountain versus valley.}
Mountain DAGs have the smallest number of nodes in the top layers and increasing numbers of nodes in subsequent layers, while valley DAGs have the largest number of nodes in the first layers with increasing numbers of nodes in subsequent layers. 

In this experiment, we consider mountain and valley DAGs with three layers and $498$ nodes in total. For mountain DAGs, there are $83, 166, 249$ nodes on the three subsequent layers. Each node in a layer is randomly assigned to one parent in its higher neighborhood layer. For valley DAGs, there are $249, 166, 83$ nodes on the three subsequent layers. Each node in a layer is randomly assigned to two parents in its higher neighborhood layer. 

\begin{figure}[h!] 
\centering
    \includegraphics[width=.45\linewidth]{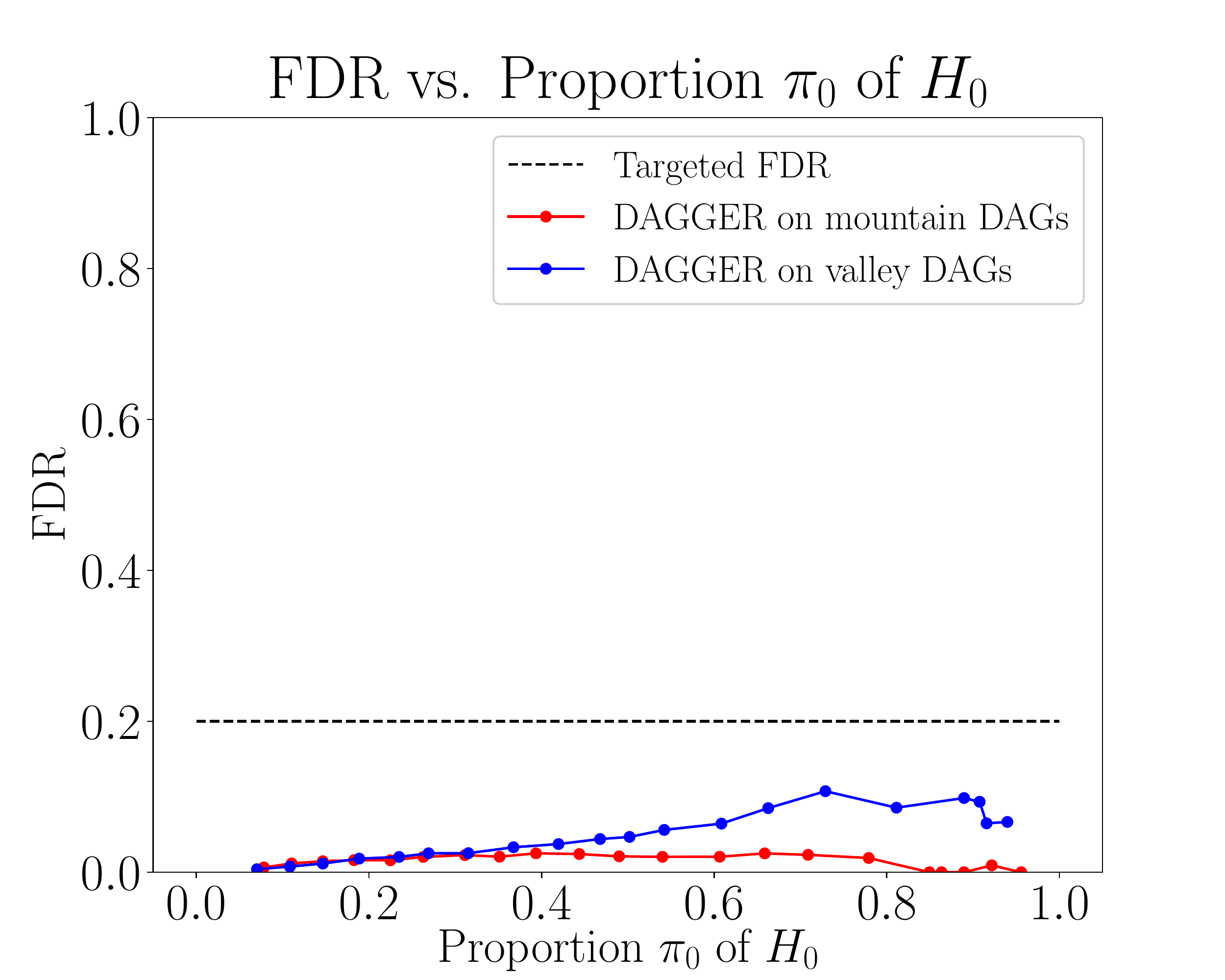}\quad 
    \includegraphics[width=.45\linewidth]{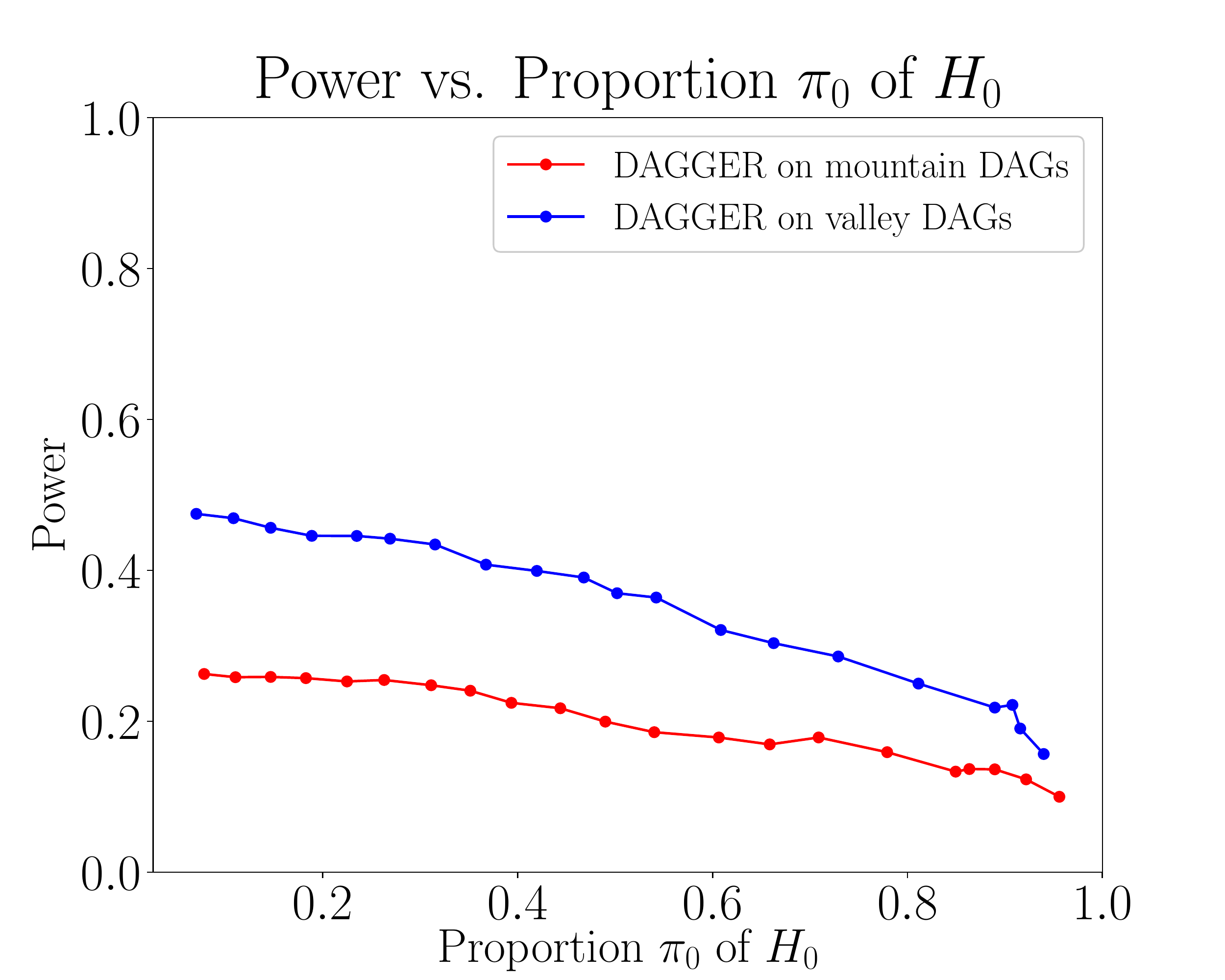} \quad
    \includegraphics[width=.45\linewidth]{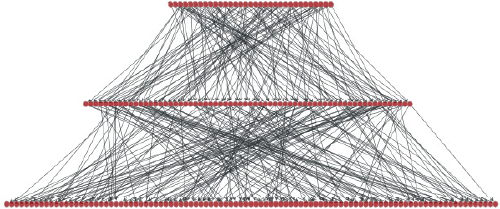}\quad 
    \includegraphics[width=.45\linewidth]{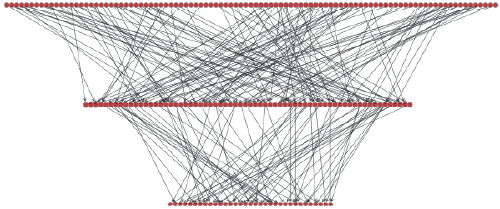}  
  \caption{Plots of the achieved FDR and power of \DAGGER versus the proportion of nulls on mountain and valley graphs respectively over $100$ repetitions, with target FDR $\alpha=0.2$, number of nodes $n=498$, signal strength $\mu=2$.}
\label{fig:mountain-valley}
\end{figure} 

Figure \ref{fig:mountain-valley} shows plots of the achieved FDR and power of \DAGGER versus the proportion of nulls on mountain and valley graphs. We set the target FDR $\alpha=0.2$ and the signal strength of non-nulls to be $\mu=2$. \DAGGER achieves a higher power on valley DAGs. Intuitively, there are smaller number of descendants for each incorrectly accepted non-nulls in valley DAGs, among which all non-nulls are ignored.

\end{document}